\pdfoutput=1
\documentclass[12pt]{article}

\usepackage{amsmath,amssymb}
\usepackage{amsthm}
\usepackage{graphicx}
\usepackage[breaklinks=true,linktocpage=true]{hyperref}
\usepackage{ascmac}
\usepackage{cite}
\usepackage{xcolor}
\numberwithin{equation}{section}
\newcommand{\fr}{\frac}

\newcommand{\er}[1]{Eq.~\eqref{#1}}

\newcommand{\hc}{\text{h.c.}}

\newcommand{\hph}{\hphantom}
\renewcommand{\b}{\bar}
\renewcommand{\d}{\dot}
\newcommand{\pd}[2]{\frac{\partial{#1}}{\partial{#2}}}

\newcommand{\sr}{\sqrt}

\newcommand{\df}{\dfrac}

\newcommand{\der}{\partial}

\renewcommand{\(}{\left(}
\renewcommand{\)}{\right)}

\newcommand{\dg}{\dagger}

\newcommand{\bmx}{\left(\begin{matrix}}
\newcommand{\emx}{\end{matrix}\right)}

\usepackage[top=2.828cm,bottom=2.828cm, left=3cm,right=3cm]{geometry}
\begin{document}
\begin{titlepage}
\hfill 
\vspace{-1em}
\def\thefootnote{\fnsymbol{footnote}}%
   \def\@makefnmark{\hbox
       to\z@{$\m@th^{\@thefnmark}$\hss}}%
 \vspace{3em}
 \begin{center}%
  {\Large 
Topological couplings in higher derivative extensions 
of supersymmetric three-form gauge theories
  \par
   }%
 \vspace{1.5em}
  {\large
Muneto Nitta\footnote{nitta@phys-h.keio.ac.jp}
and 
Ryo Yokokura\footnote{ryokokur@keio.jp}
   \par
}%
 \vspace{1.5em}
{\small \it
Department of Physics \& 
Research and Education Center for Natural Sciences,
Keio University, Hiyoshi 4-1-1, Yokohama, Kanagawa 223-8521, Japan
 \par
}
\vspace{1.5em} 
   {\large 
    }%
 \end{center}%
 \par
\vspace{1.5em}% 
\begin{abstract}
We consider a topological coupling between 
a pseudo-scalar field and a 3-form gauge field in 
 ${\cal N}=1$ supersymmetric higher derivative 
3-form gauge theories in four spacetime dimensions.
We show that ghost/tachyon-free higher derivative Lagrangians with 
the topological coupling 
can generate various potentials for the pseudo-scalar field
by solving the equation of motion for the 3-form gauge field.
We give two examples of higher derivative Lagrangians and 
the corresponding potentials: one is a quartic order term of the 
field strength and the other is the term which can generate a 
cosine-type potential of the pseudo-scalar field.
\end{abstract}
\end{titlepage}
 \setcounter{footnote}{0}%
\def\thefootnote{$*$\arabic{footnote}}%
   \def\@makefnmark{\hbox
       to\z@{$\m@th^{\@thefnmark}$\hss}}% 
\tableofcontents
\section{Introduction}
3-form gauge theories in four dimensions (4D)
 have been investigated in various contexts.
In quantum chromodynamics (QCD), 
3-form gauge theories have been applied to
confinement~\cite{Aurilia:1977jz,Aurilia:1978dw}, 
 an effective description of 
the Chern--Simons 3-form~\cite{Luscher:1978rn,Aurilia:1978dw}, 
$U(1)$ problem~\cite{Aurilia:1980jz,Hata:1980hn}, 
and strong CP problem~\cite{Dvali:2005an,Dvali:2005zk}.
In cosmology, the 3-form gauge theories have been studied in the
cosmological constant problem~\cite{Aurilia:1980xj,Hawking:1984hk,Brown:1987dd,Brown:1988kg,Duff:1989ah,Duncan:1989ug},
quintessence~\cite{Kaloper:2008qs,DAmico:2018mnx}, 
and inflationary models~\cite{Kaloper:2008fb,Kaloper:2011jz,Marchesano:2014mla,Kaloper:2014zba,Kaloper:2016fbr,DAmico:2017cda}.
The 3-form gauge theories are also related to 
 condensed matter physics, e.g., superconductor~\cite{Ansoldi:1995by}
and quantum Hall effect~\cite{Hasebe:2014nia}.

A supersymmetric (SUSY) extension of the 3-form gauge theories 
was introduced in Ref.~\cite{Gates:1980ay},
and then it was embedded into supergravity 
(SUGRA)~\cite{Gates:1980az,Buchbinder:1988tj,Binetruy:1996xw}
(see Refs.~\cite{Gates:1983nr,Buchbinder:1998qv,Binetruy:2000zx} as a
 review).
The SUSY 3-form gauge theories were also extensively studied 
with various applications: 
gaugino condensation in SUSY Yang--Mills 
theories~\cite{Binetruy:1995hq,Binetruy:1995ta,Dvali:1999pk}, 
coupling with a membrane~\cite{Ovrut:1997ur,Kuzenko:2017vil,Bandos:2018gjp},
alternative formulation of old-minimal 
SUGRA~\cite{Ovrut:1997ur,Kuzenko:2005wh,Farakos:2016hly,Farakos:2017jme},
St\"uckelberg coupling~\cite{Groh:2012tf,Hartong:2009az,Becker:2016xgv,Aoki:2016rfz}, 
topological coupling~\cite{Dudas:2014pva,Becker:2016xgv,Yokokura:2016xcf},
complex 3-form gauge theories~\cite{Farakos:2017jme,Kuzenko:2017vil,Bandos:2018gjp},
extended SUSY~\cite{Cribiori:2018jjh},
the cosmological constant problem~\cite{Farakos:2016hly},
SUSY breaking~\cite{Farakos:2016hly,Buchbinder:2017vnb,Kuzenko:2017oni},
and inflationary models~\cite{Dudas:2014pva,Yamada:2018nsk}.

One of the virtues of the 3-form gauge theories is
that the 3-form gauge field can generate a potential for a 
pseudo-scalar field with a topological coupling between
the 3-form and the pseudo-scalar 
field~\cite{Aurilia:1980jz,Curtright:1980yj,Dvali:2005an,Dvali:2005zk,Kaloper:2008qs,Kaloper:2008fb}.
This mechanism was applied to effective theories of
QCD~\cite{Aurilia:1980jz,Dvali:2005an,Dvali:2005zk,Kaloper:2017fsa}, 
cosmology~\cite{Kaloper:2008qs,Kaloper:2008fb,Kaloper:2011jz,Kaloper:2014zba,Marchesano:2014mla,Franco:2014hsa,Kaloper:2016fbr,DAmico:2017cda,DAmico:2018mnx},
and string effective theories~\cite{Bielleman:2015ina,Valenzuela:2016yny,Montero:2017yja}.
The pseudo-scalar field is often assumed 
to have a shift symmetry like a Goldstone boson or an axion, 
but a potential term generally break the shift symmetry of the 
pseudo-scalar field.
It was pointed out that the topological coupling preserves the shift symmetry of Lagrangians, 
and the shift symmetry is spontaneously 
broken~\cite{Kaloper:2008qs,Kaloper:2008fb}.
A merit of this mechanism is that the potentials can be obtained by only assuming the symmetries of infrared theories, but not necessarily depend on details of ultraviolet models.  
This mechanism can also be considered in SUSY theories.
In the SUSY theories, the field strength of the 3-form gauge field
can be embedded into an auxiliary field of a chiral 
superfield given by a real superfield~\cite{Gates:1980ay}.
The topological coupling is given by 
choosing appropriate 
superpotential~\cite{Dudas:2014pva,Becker:2016xgv,Yokokura:2016xcf}.
This coupling was applied to SUGRA inflationary
models~\cite{Dudas:2014pva,Yamada:2018nsk}.

In the above applications, 
the 3-form gauge theories are often  regarded 
as infrared (low energy or long range) effective theories.
Since effective theories inevitably include nonrenormalizable 
interactions which generally
depend on an ultraviolet (high energy or 
short range) cutoff parameter, it is natural to consider 
such nonrenormalizable terms in the 3-form gauge theories.
The nonrenormalizable terms may be given by functions of 
the field strength of the 3-form gauge field rather than 
the 3-form gauge field itself because 
Lagrangians should be gauge invariant.
Since the field strength is given by a 
spacetime derivative of the gauge field, 
the nonrenormalizable terms given by the field strength are regarded as higher derivative terms.

The higher derivative terms with the topological 
coupling in the 3-form gauge theories
can generate various potentials for the 
pseudo-scalar field while preserving the shift symmetry for 
the pseudo-scalar field~\cite{Dvali:2005an,Dvali:2005zk,Kaloper:2011jz,DAmico:2017cda}.
Indeed, a cosine-type potential for the pseudo-scalar field 
can be generated by choosing appropriate higher 
derivative terms~\cite{Dvali:2005an}.
This is in contrast to the case of 
 the quadratic (canonical) kinetic term for which
only the mass term (quadratic potential) for the pseudo-scalar field can be generated.
This mechanism has been applied to 
the strong CP problem~\cite{Dvali:2005an,Dvali:2005zk} 
and inflationary models~\cite{Kaloper:2011jz,DAmico:2017cda}. 
For SUSY 3-form gauge theories, generating the potentials by
the topological coupling and the higher derivative terms
has been described in a quartic order of the field 
strength~\cite{Dudas:2014pva}.
However, the general form of the potentials for the pseudo-scalar field has not been understood in our knowledge.
The purpose of this paper is to 
investigate this general form of the potentials 
generated by the topological coupling and the higher derivative terms 
in SUSY 3-form gauge theories.

In the 3-form gauge theories, however, 
some higher derivative terms may lead to a tachyon
instability when derivatives of the field strength are included
in addition to the canonical kinetic term~\cite{Nitta:2018yzb}.
This instability is different from that of scalar field theories.
For scalar field theories, 
 derivative terms on a scalar field higher than the first order 
may give a so-called Ostrogradsky's 
ghost instability~\cite{Ostrogradsky:1850fid,Woodard:2006nt}.
While for the 3-form gauge theories, 
 the derivative terms of the field strength of the 3-form gauge 
field~\cite{Klinkhamer:2016jrt,Klinkhamer:2016zzh,Klinkhamer:2016lgk} 
do not cause such ghosts in general~\cite{Klinkhamer:2017nfb} 
but may cause a tachyon.
Note that the tachyon can also be a dynamical ghost depending on the parameter of the higher derivative term. 
This ghost is different from the Ostrogradsky's ghost.

One of the sufficient conditions for the ghost/tachyon-free Lagrangians 
is that the Lagrangians should have an arbitrary function of
 the field strength 
but do not consist of derivative terms 
of the field strength of a 3-form gauge field~\cite{Nitta:2018yzb}.
Previously known models in Refs.~\cite{Dvali:2005an,Dvali:2005zk,Kaloper:2011jz,DAmico:2017cda} fall into this class.
This condition is not a necessary condition: 
if the canonical kinetic term 
is absent or has the wrong sign~\cite{Klinkhamer:2017nfb}, 
there is no tachyon instability.
In SUSY 3-form gauge theories, the ghost/tachyon-free 
Lagrangians were considered in the quartic order of the field 
strength~\cite{Dudas:2014pva}.
Later, 
the Lagrangians with an arbitrary order of the field strength were 
established~\cite{Nitta:2018yzb}.
Note that these SUSY Lagrangians are based on 
ghost-free Lagrangians for chiral superfields formulated 
in Refs.~\cite{Khoury:2010gb,Khoury:2011da,Koehn:2012te,Nitta:2014pwa},
which have been applied in 
many contexts in SUSY effective field theories~\cite{Buchbinder:1994iw,Buchbinder:1994xq,Banin:2006db,Kuzenko:2014ypa,Koehn:2012ar,Farakos:2012qu,Queiruga:2016yzd,Sasaki:2012ka,Aoki:2014pna,Aoki:2015eba,Adam:2013awa,Adam:2011hj,Nitta:2015uba,Bolognesi:2014ova,Queiruga:2016jqu,Queiruga:2018nph,Gudnason:2015ryh,Gudnason:2016iex,Queiruga:2015xka,Queiruga:2017blc,Eto:2012qda,Nitta:2014fca,Nitta:2017yuf,Nitta:2017mgk,Farakos:2018sgq}\footnote{
For vector superfields,
ghost-free higher derivative Lagrangians were considered in 
Refs.~\cite{Cecotti:1986jy,Bagger:1996wp,Kuzenko:2000uh,Kuzenko:2002vk,Antoniadis:2007xc,Farakos:2012qu,Dudas:2015vka}.
The ghost-free higher derivative Lagrangian with an arbitrary order of the 
field strength was formulated in Ref.~\cite{Fujimori:2017kyi}
and applied in Refs.~\cite{Cribiori:2017laj,Aldabergenov:2018nzd,Kuzenko:2018jlz,Aldabergenov:2017hvp,Abe:2018plc}}.

In this paper, we show a general framework and concrete examples 
of the generation of such the potentials by using 
the topological coupling and 
the higher derivative terms in bosonic and ${\cal N}=1$ 
SUSY 3-form gauge fields.
For the bosonic 3-form gauge theories,
we show a relation between the ghost/tachyon-free
higher derivative terms 
and generated potentials.
While the relation has been known by using dual formulations 
of the 3-form gauge theories in Ref.~\cite{Dvali:2005an},
we show a direct derivation which 
is given in terms of the 3-form gauge field itself.

For the SUSY 3-form gauge theories, we  
use ghost/tachyon-free Lagrangians of the 3-form gauge field 
with an arbitrary order of the field strength~\cite{Nitta:2018yzb}
in order to avoid the instabilities explained above.
First, we give a bosonic part of the SUSY Lagrangian 
with the topological coupling and an arbitrary order of the field strength by using the ghost/tachyon-free Lagrangian of the 3-form gauge field.
Second, we show a general expression of 
the potentials of the pseudo-scalar field generated by 
the topological coupling.
Finally, we give two examples of the higher derivative terms and 
the generated potentials.
One is a quartic order of the field strength. 
This quartic higher derivative term 
was described in Ref.~\cite{Dudas:2014pva}, 
and we derive the generated potential explicitly.
Another is a cosine-type potential.
We show that such the potential can also be 
obtained by choosing an appropriate SUSY higher derivative terms.

This paper is organized as follows.
In Sec.~\ref{b}, we first review the potential generation mechanism in 
bosonic 3-form gauge theories.
We then show a direct derivation 
of the generated potential~\cite{Dvali:2005an}
in terms of the 3-form gauge field itself.
In Sec.~\ref{s}, we discuss the SUSY extension of the potential 
generation mechanism in the 3-form gauge theories.
In Sec.~\ref{sum}, we summarize this paper.
We use the notations and conventions of Ref.~\cite{Wess:1992cp}.

\section{3-form gauge theories with a topological coupling}
\label{b}
In this section, we first review the role of the topological coupling 
with a pseudo-scalar field 
to generate a mass term (potential) of the pseudo-scalar field
in 3-form gauge theories~\cite{Aurilia:1980jz,Dvali:2005an,Dvali:2005zk,Kaloper:2008qs,Kaloper:2008fb}.
We then show the relation between the ghost/tachyon-free higher derivative terms and the generated potential in terms of the 3-form gauge field
itself rather than a dual formulation of the 3-form gauge theories. 
A virtue of the 3-form is a mass (potential) 
generation of a shift symmetric pseudo-scalar field.

\subsection{3-form gauge field}
\label{bc}
A 3-form gauge field $C_{mnp}$
is a third-rank anti-symmetric tensor \cite{Aurilia:1977jz}.
The gauge tranformation is given by
\begin{equation}
 \delta_3 C_{mnp} = \der_m \lambda_{np} +\der_n \lambda_{pm} + \der_p \lambda_{mn}, 
\end{equation}
where $\delta_3$ denotes an infinitesimal gauge transformation of 
3-form gauge field,
and $\lambda_{mn}$ is a second-rank 
anti-symmetric tensor parameter. 
To describe the kinetic term for the 3-form gauge field, 
we introduce the field strength of the 3-form gauge field as
\begin{equation}
F_{mnpq} = \der_m C_{npq} - \der_n C_{mpq} + 
\der_p C_{mnq} - \der_q C_{mnp}. 
\end{equation}
It is convenient to define the Hodge dual of the field strength
$F:= \fr{1}{4!} \epsilon^{mnpq} F_{mnpq}$, where
$\epsilon^{mnpq}$ is the totally anti-symmetric tensor 
with the normalization $\epsilon^{0123} = +1$.

Then the quadratic (canonical) kinetic term for the 3-form gauge field
is given by
\begin{equation}
 {\cal L}_{\rm kin.} 
= -\fr{1}{2\cdot 4!} F^{mnpq}F_{mnpq} 
+\fr{1}{3!} \der_m (C^{mnp}F_{mnpq})
= +\fr{1}{2} F^2 
-\fr{1}{3!} \der_m (\epsilon_{mnpq}C^{mnp}F)
\label{181026.1752}
\end{equation}
with the following gauge invariant boundary condition for the 
3-form gauge field:
\begin{equation}
 F|_{\rm bound.} = -f_0,
\label{181026.1905}
\end{equation}
where the symbol $|_{\rm bound.}$ denotes the value at the boundary,
$f_0$ is a real constant, and the minus sign is 
just a convention.
The term $\fr{1}{3!} \der_m (C^{mnp}F_{mnpq})$ 
in \er{181026.1752} is the boundary term for the 3-form 
gauge field, which is needed for the vanishing of the 
functional variation of the 3-form gauge field at the boundary.
If this boundary term were absent, the functional variation 
of the 3-form gauge field would not vanish, and the equation of motion (EOM) for the 
3-form gauge field and the energy-momentum tensor would be 
inconsistent~\cite{Duff:1989ah,Duncan:1989ug}.

\subsection{Topological coupling with a pseudo-scalar field}
\label{btc}
In this subsection, we introduce a topological coupling between the 3-form 
gauge field and a pseudo-scalar field~\cite{Aurilia:1980jz,Dvali:2005an,Dvali:2005zk,Kaloper:2008qs,Kaloper:2008fb}.
This coupling can be applied to the generation of a 
potential compatible with a shift symmetry for the pseudo-scalar field~\cite{Kaloper:2008qs,Kaloper:2008fb}.
In general, the shift symmetry is broken by 
 a potential term for the pseudo-scalar 
field.
However, there is a mechanism of the
generation of the potential which is compatible 
with the shift symmetry of the pseudo-scalar field
if we consider a 3-form gauge field:
the potential can be generated by
a topological coupling between 
the 3-form gauge field and the pseudo-scalar field $\phi$
of the form $\phi \epsilon^{mnpq} F_{mnpq}$.

The Lagrangian is given by
\begin{equation}
\begin{split}
 {\cal L}_\text{top.}
&= 
-\df{1}{2\cdot 4!} F^{mnpq} F_{mnpq}
-\df{1}{2} \der^m \phi \der_m \phi
+\df{1}{4!} m \phi \epsilon^{mnpq} F_{mnpq}
\\
&\quad 
+\df{1}{3!}\der^m(C^{npq}F_{mnpq} - m \phi \epsilon_{mnpq}C^{npq})
\end{split}
\label{180719.1854}
\end{equation} 
with the boundary condition for the pseudo-scalar field
\begin{equation}
 \phi|_{\rm bound.} = \phi_0
\label{181026.1906}
\end{equation}
and the one for the field strength in \er{181026.1905}.
In \er{180719.1854},
$m$ is a mass-dimension one 
real parameter which will be  
a mass of the pseudo-scalar field $\phi$.
The boundary term 
$+\tfrac{1}{3!}\der^m(C^{npq}F_{mnpq} - m \phi \epsilon_{mnpq}C^{npq})$
is determined by the requirement that 
the variation of the 3-form at the boundary 
should vanish.

The Lagrangian has a shift symmetry of the pseudo-scalar field
\begin{equation}
 \phi \to \phi +k,
\end{equation}
where  $k$ is a real constant.
Under the shift of the pseudo-scalar field, the Lagrangian 
is varied as
\begin{equation}
\begin{split}
 {\cal L}_\text{top.}
&\to 
 {\cal L}_\text{top.}
+
\df{1}{4!} m  k \epsilon^{mnpq} F_{mnpq}
-\df{1}{3!} m k \der^m (\epsilon_{mnpq} C^{npq})
\\
&=
 {\cal L}_\text{top.}
+
\df{1}{4!} m k \epsilon^{mnpq} F_{mnpq}
-\df{1}{4!} m k \epsilon_{mnpq} F^{mnpq}
=0. 
\end{split}
\end{equation}
Thus, the shift of the bulk term is 
canceled by that of the 
boundary term.

The generation of the mass term can be seen by solving EOM of
the 3-form gauge field.
The EOM of the 3-form gauge field is 
\begin{equation}
 0 = \der^m F_{mnpq} 
- 
\epsilon_{mnpq} m \der^m \phi. 
\end{equation}
Therefore, we can solve the EOM as 
\begin{equation}
 F_{mnpq} = \epsilon_{mnpq}(m\phi +c),
\end{equation}
where $c$ is a constant determined by the boundary conditions
 for $F$ and $\phi$ as $c = f_0 - m\phi_0$.
Substituting this solution into the Lagrangian in \er{180719.1854},
we obtain 
\begin{equation}
\begin{split}
 {\cal L}_\text{top.}
&
= 
+\df{1}{2} (m\phi + c)^2
-\df{1}{2} \der^m \phi \der_m \phi
- m \phi (m\phi + c)
+\df{1}{3!}c \der^m(\epsilon_{mnpq}C^{npq})
\\
&= 
-\df{1}{2} \der^m \phi \der_m \phi
-\df{1}{2} (m\phi + c)^2.
\end{split}
\label{180719.1854-2}
\end{equation} 
Therefore, the topological 
coupling gives us the mass of the pseudo-scalar field.
\subsection{Higher derivative term and topological coupling}
\label{bth}
We have seen that the Lagrangian with the canonical kinetic term 
and the topological coupling can generate the mass term for the 
pseudo-scalar field.
This mechanism can be generalized to 
more complicated potentials than the mass term.
Such potentials can be generated by introducing 
higher derivative interactions for the 3-form gauge field 
as shown below.
We will consider a class of the higher derivative terms 
which do not cause 
either tachyon nor ghost instability.
The higher derivative terms for the 3-form gauge fields
often cause tachyon and/or ghost instability, 
although such the instability is absent if there are 
no derivative terms on the field strength.

A ghost/tachyon-free higher derivative Lagrangian 
with the topological coupling can be written as
\begin{equation}
 {\cal L}_{\rm top., HD}
 = G(F) 
-\df{1}{2} \der^m \phi \der_m \phi 
+m \phi F
-\df{1}{3!}\epsilon^{mnpq} \der_m ( G'(F) C_{npq} + m\phi C_{npq}).
\label{180725.2141}
\end{equation}
Here, $G$ is an arbitrary function of the field strength~\cite{Dvali:2005an,Dvali:2005zk}.
The canonical kinetic term can be included into $G$ as
$G(F) = +\fr{1}{2} F^2\cdots$.
The term 
$-\fr{1}{3!}\epsilon^{mnpq} \der_m ( G'(F) C_{npq} + m\phi C_{npq})$ 
is the boundary term for the higher derivative term
and the topological term.
The boundary term for the higher derivative term has been 
determined by requiring the vanishing of the functional variation 
of the 3-form at the boundary~\cite{Nitta:2018yzb}.
We assume the same boundary conditions for 
the field strength $F$ and the pseudo-scalar field $\phi$
as \er{181026.1905} and \er{181026.1906}, respectively.

The potential for the pseudo-scalar field can be obtained by 
solving the EOM for the 3-form gauge field as in 
the previous case.
The EOM for the 3-form gauge field is
\begin{equation}
  0 = \der_m G'(F)  + m\der_m \phi. 
\end{equation}
This can be solved as
\begin{equation}
 - G'(F) =  m\phi + c,
\label{181011.2014}
\end{equation}
where $c$ is a constant determined by the boundary condition for 
the field strength and the pseudo-scalar field.
By this equation, the field strength $F$ 
can be implicitly solved in terms of the pseudo-scalar field $\phi$:
$F = F(\phi)$.
Then, the Lagrangian becomes
\begin{equation}
 {\cal L}_{\rm top., HD}
 = 
-\df{1}{2} \der^m \phi \der_m \phi
+
G(F) -F G'(F).
\label{180725.2141-2}
\end{equation}
The term $G(F) -F G'(F)$ is now the potential for the pseudo-scalar
field.
If we choose the function $G(F)$ as the canonical kinetic term as
$G(F) = +\fr{1}{2} F^2$, we can obtain the mass term, 
since the solution to the EOM and 
the potential term $G(F) -FG'(F) $ are $F +m\phi = -c$ and 
$-\fr{1}{2}(m \phi + c)^2$, respectively.

Generally, we can obtain an arbitrary potential 
for the pseudo-scalar field $V(\phi)$ by using this topological 
coupling~\cite{Dvali:2005an}. 
The potential $V$ can be given 
in terms of the 3-form gauge field as
\begin{equation}
V(\phi) = -G(F) +FG'(F).
\label{181024.0427}
\end{equation}
The derivative of the both hand sides by $\phi$ is 
$V'(\phi) = F G''(F) \pd{F}{\phi}$, and the derivative of 
\er{181011.2014} by $\phi$ leads to 
\begin{equation}
V'(\phi) = - m F.
 \label{181024.0444}
\end{equation}
Thus, the relation between the higher derivative term and the 
potential is 
\begin{equation}
V'(\phi) = -m (G')^{-1} (-m\phi-c).
\label{181011.2022}
\end{equation}
We should comment on the comparison with 
the previous work in Ref.~\cite{Dvali:2005an}.
In the procedure in Ref.~\cite{Dvali:2005an},
the dual description of the 3-form was 
applied to derive \er{181011.2022}.
Now, we have shown that the mechanism of 
the potential generation is directly obtained in terms of 
the 3-form gauge field itself.

As a concrete example, we review the generation of 
a cosine potential for the pseudo-scalar field 
given in Ref.~\cite{Dvali:2005an}.
The Lagrangian is given by
\begin{equation}
\begin{split}
 {\cal L}_{\rm cos}
&= 
 M^2 F \arcsin \df{F}{M^2} 
+
M^4 \sr{1-\df{F^2}{M^4}}
-\df{1}{2}\der^m \phi \der_m\phi 
+ m \phi F
\\
&
\quad
-\df{1}{3!}\epsilon^{mnpq} \der_m \( C_{npq} M^2 \arcsin\(\fr{F}{M^2}\) + m\phi C_{npq}\)
, 
\end{split}
\label{108725.2154}
\end{equation}
where $M$ is a scale parameter with mass dimension one.
Here, the higher derivative term has been chosen so that 
the relation between the potential and the higher derivative 
term in \er{181011.2022} is satisfied:
\begin{equation}
 G(F) =  M^2 F \arcsin \df{F}{M^2} 
+
M^4 \sr{1-\df{F^2}{M^4}}.
\label{181025.0814}
\end{equation}

The EOM for the 3-form gauge field is 
\begin{equation}
 \df{1}{\sr{1-\df{F^2}{M^4}}}\der_m F 
+ m \der_m \phi  = 0.
\end{equation}
This can be solved as
\begin{equation}
 m \phi +c = - M^2 \arcsin\(\df{F}{M^2}\),
\end{equation}
where $c$ is a constant.
Thus, we can express $F$ in terms of $\phi$:
\begin{equation}
 F = - M^2 \sin\(\df{m\phi +c}{M^2}\) .
\end{equation}
Substituting the solution into
\er{108725.2154}, we obtain
\begin{equation}
 {\cal L}_{\rm cos}
 = 
-\df{1}{2}\der^m \phi \der_m \phi
+M^4 \cos \(\df{m\phi+c}{M^2}\).
\end{equation}

\section{Higher derivative terms and scalar potentials 
in SUSY 3-form gauge theories}
\label{s}
In this section, we consider an ${\cal N}=1$ 
SUSY extension of the generation of the 
potentials of a pseudo-scalar field by the topological 
coupling and higher derivative terms for the 3-form gauge field.
In this paper, we use superspace in order to formulate 
manifestly SUSY theories.
The superspace is spanned by the 
bosonic spacetime coordinates $(x^m)$ 
and the fermionic coordinates given by Grassmann numbers
$(\theta^\alpha,\b\theta_{\d\alpha})$,
where undotted and dotted Greek letters beginning with 
$\alpha, \beta,...$ and $\d\alpha,\d\beta,...$ 
denote undotted and dotted spinors, respectively. 
Thus, the coordinates of the superspace are 
denoted as $(x^m, \theta^\alpha,\b\theta_{\d\alpha})$.
We use the notation and convention of the Ref.~\cite{Wess:1992cp}.

\subsection{3-form gauge field in superspace}
\label{sc}
Here, we briefly review SUSY 3-form gauge 
theories~\cite{Gates:1980ay,Groh:2012tf,Farakos:2017jme}.
In the superspace, a 
3-form gauge field $C_{mnp}$ is embedded into a real 
superfield $X$ satisfying $X^\dg = X$ as follows:
\begin{equation}
 C_{mnp} = 
\fr{\sr{2}}{8}\epsilon_{mnpq} (\b\sigma^q)^{\d\alpha\alpha}
[D_\alpha, \b{D}_{\d\alpha}] X|.
\end{equation}
Here, 
$(\b\sigma^q)^{\d\alpha\alpha}$ denotes 4D Pauli matrices,
 $D_\alpha $ and $\b{D}_{\d\alpha}$ are 
super-covariant spinor derivatives, and
the vertical bar ``$|$'' denotes the $\theta = \b\theta =0$
projection in the superspace. 
Hereafter, we will call 
$X$ ``3-form prepotential''~\cite{Gates:1983nr}.
The gauge transformation of the 3-form 
prepotential in the superspace is given by 
a chiral superfield parameter $\Upsilon_\alpha$ satisfying 
$\b{D}_{\d\alpha} \Upsilon_\alpha =0$ as
\begin{equation}
 \delta_{3,\rm SUSY} X
 = 
\fr{1}{2i}(D^\alpha \Upsilon_\alpha 
- \b{D}_{\d\alpha} \b\Upsilon^{\d\alpha}).
\label{181012.2111}
\end{equation}
The field strength of the 3-form gauge field is embedded 
into a chiral superfield $Y$, which is related to 
the original real superfield $X$ as follows:
\begin{equation}
 Y = -\df{1}{4} \b{D}^2 X,
\end{equation}
\begin{equation}
 F_{mnpq} = \epsilon_{mnpq} \fr{\sr{2}i}{8} (D^2 Y-\b{D}^2\b{Y})|.
\end{equation}
Note that $Y$ is invariant under the gauge transformation in 
\er{181012.2111}.
In the SUSY 3-form gauge theories,
there are superpartners of the 3-form gauge field,
and they are also embedded into the chiral superfield $Y$.
The superpartners are a complex scalar field $y = Y|$,
a Weyl fermion $\chi_\alpha = \fr{1}{\sr{2}}D_\alpha Y|$,
and a real auxiliary field 
$H = -\fr{\sr{2}}{8}(D^2 Y + \b{D}^2 \b{Y})|$.
It will be useful to define the combination of the field strength
and the auxiliary field $H$ as
\begin{equation}
 {\cal F} = -\fr{1}{4}D^2 Y |  = \fr{1}{\sr{2}}(H - iF).
\end{equation}
In SUSY 3-form gauge theories, 
we assume the boundary conditions for the 3-form prepotential.
The boundary condition for the 3-form gauge field is 
\begin{equation}
 \delta Y |_{\rm bound.}  
= -\fr{1}{4}(\b{D}^2 \delta X) |_{\rm bound.}
= 0,
\label{181028.1714}
\end{equation}
where $\delta$ denotes the functional variation of fields.
This boundary condition leads to 
$
\delta y|_{\rm bound.}  = \delta \chi_\alpha |_{\rm bound.} = 
\delta H |_{\rm bound.} = \delta F |_{\rm bound.} =0.$
We further assume the same boundary condition for the 
field strength $F$ as
\begin{equation}
 F|_{\rm bound.}  = -f_0,
\label{181028.1715}
\end{equation}
which is consistent with the boundary condition in \er{181028.1714}.

The Lagrangian with the canonical kinetic term of the 
3-form gauge field can be written by 
\begin{equation}
\begin{split}
 {\cal L}_{\rm kin.,SUSY}
& 
= 
-\fr{1}{8}\int d^2\theta \b{D}^2  Y\b{Y}
+
\fr{i}{8} 
\(
\int d^2 \theta \b{D}^2
-\int d^2\b\theta D^2
\) T_I X
+\hc,
\end{split}
\end{equation}
where $\int d^2 \theta = -\fr{1}{4} D^2 |$ 
is the so-called F-type integration.
We use $-\fr{1}{8} (\int d^2\theta \b{D}^2 + \int d^2\b\theta D^2)$ 
for the so-called D-type integration in stead of conventional 
$\int d^4\theta $ to fix the definition of the D-type integration.
The first term leads to 
the canonical kinetic term of the 3-form gauge field,
while the second term gives us the boundary term for the 
first term.
The superfield $T_I$ is the imaginary part of the chiral 
superfield $T$: $T_I =\fr{1}{2i}(T-\b{T})$,
and $T$ is defined by 
\begin{equation}
 T = -\fr{1}{4} \b{D}^2 \b{Y}.
\end{equation}
The chiral superfield $T$ is chosen by the requirement that 
 the functional variation of the prepotential $X$ should 
vanish at the boundary.

The bosonic part of the 
Lagrangian can be expressed in terms of the component fields
as follows:
\begin{equation}
 {\cal L}_{\rm kin.,SUSY}
 = -\der^m y \der_m \b{y}
+\fr{1}{2} F^2 + \fr{1}{2} H^2
-\fr{1}{3!}\der_m (\epsilon^{mnpq} F C_{npq} ),
\end{equation}
where we have used the Wess--Zumino (WZ) gauge for the 
3-form prepotential:
\begin{equation}
 X| =0 ,
\quad
D_\alpha X| =0,
\quad
\b{D}_{\d\alpha} X |= 0.
\end{equation}
Note that we hereafter omit fermions in the Lagrangians
for the component fields, which will be 
irrelevant to the discussion of the generation of the 
potentials.
\subsection{Scalar potential for a pseudo-scalar field}
\label{stc}
Here, we review the generation of the scalar potential for the 
pseudo-scalar field
by topological coupling and the canonical kinetic term for the 
3-form gauge field in SUSY field theories~\cite{Dudas:2014pva}.
The pseudo-scalar field $\phi$ is embedded into a chiral 
superfield $\Phi$:
\begin{equation}
 \phi = \fr{\sr{2}}{2i} (\Phi - \b\Phi)|.
\end{equation}
The shift transformation of the pseudo-scalar field is 
given by 
\begin{equation}
 \Phi \to \Phi + ik,
\label{181013.1907}
\end{equation}
where $ k $ is a real constant.
In the SUSY theories, there are superpartners of the pseudo-scalar 
field.
The superpartners of the pseudo-scalar field are 
a real scalar field $s$, 
a Weyl fermion $\psi_\alpha$,
and a complex auxiliary field $F_\Phi$.
They are given by
\begin{equation}
 s = \fr{\sr{2}}{2} (\Phi+ \b\Phi)|,
\quad
\psi_\alpha = \fr{1}{\sr{2}} D_\alpha \Phi|, 
\quad
F_\Phi = -\fr{1}{4}D^2 \Phi|.
\end{equation}
It will be convenient to define the following complex 
scalar field
\begin{equation}
 \varphi = \Phi| =\fr{1}{\sr{2}} (s + i\phi),
\end{equation}
which we will use in section~\ref{sth}.
We assume the same boundary condition for the pseudo-scalar field
as \er{181026.1906}:
\begin{equation}
 \phi |_{\rm bound.} = \phi_0.
\end{equation}
Note that the boundary conditions for the other fields in $\Phi$
will not be used in the following discussion, therefore
we do not show them explicitly.

The topological coupling between 
the pseudo-scalar field and the 3-form gauge field 
can be introduced by the superpotential $m \Phi Y$.
The total Lagrangian is given by 
\begin{equation}
\begin{split}
 {\cal L}_{\rm top.,SUSY}
& = -\fr{1}{8}\int d^2\theta \b{D}^2 
\(\fr{1}{2}(\Phi+\b\Phi)^2 + Y\b{Y}\)
+m \int d^2\theta \Phi Y 
\\
&
\quad
+
\fr{i}{8} 
\(
\int d^2 \theta \b{D}^2
-\int d^2\b\theta D^2
\) T_{I, \rm top.} X
+\hc
\end{split}
\end{equation}
Here, the second line of the Lagrangian is the boundary term. 
The superfield $T_{I, \rm top.}$ in the boundary term 
is the imaginary part of the chiral superfield
$T_{\rm top.}$ given by 
\begin{equation}
 T_{\rm top.} = -\fr{1}{4} \b{D}^2 \b{Y} + m \Phi.
\end{equation}
This chiral superfield is chosen so that 
 the functional variation of the prepotential $X$ should 
vanish at the boundary.
This Lagrangian is invariant under the shift 
transformation of the chiral superfield $\Phi$ in \er{181013.1907}.
The shift transformations of the superpotential and the 
boundary term are canceled by each other, since these 
transformations are given by
\begin{equation}
m \int d^2\theta \Phi Y 
+ \hc
\to  
m \int d^2\theta \Phi Y 
+ 
m i k \int d^2\theta Y 
+ \hc
\end{equation}
and 
\begin{equation}
\begin{split}
&\fr{i}{8} 
\(
\int d^2 \theta \b{D}^2
-\int d^2\b\theta D^2
\) T_{I, \rm top.} X
+\hc
\\
&
\to
\fr{i}{8} 
\(
\int d^2 \theta \b{D}^2
-\int d^2\b\theta D^2
\) (T_{I, \rm top.} + m k) X
+\hc
\\
&
=
\fr{i}{8} 
\(
\int d^2 \theta \b{D}^2
-\int d^2\b\theta D^2
\) T_{I, \rm top.}  X
-
\fr{i}{2} m k
\(
\int d^2 \theta Y
-\int d^2\b\theta \b{Y}
\) 
+\hc
\end{split}
\end{equation}

The scalar potential for the pseudo-scalar field can be seen 
by solving the EOM for the auxiliary fields and the field strength of the 3-form gauge field.
In order to obtain the EOM, it is convenitent to 
express the bosonic part of the Lagrangian.
The bosonic part of the Lagrangian is 
\begin{equation}
\begin{split}
 {\cal L}_{\rm top., SUSY}
& = 
-\der^m y \der_m \b{y}
+\fr{1}{2} F^2 + \fr{1}{2} H^2
- \fr{1}{2} \der^m \phi \der_m \phi 
- \fr{1}{2} \der^m s \der_m s 
+ F_\Phi \b{F}_{\Phi}
\\
&
\quad
+ m (y F_\Phi + \b{y} \b{F}_\Phi + F \phi +  H s) 
-\fr{1}{3!}\der_m (\epsilon^{mnpq} (F + m\phi)  C_{npq} ),
\end{split} 
\label{181013.2038}
\end{equation}
where we have used the WZ gauge.
The EOM for the field strength and the auxiliary fields are
\begin{equation}
F +m\phi = -c, 
\quad
H = -ms, 
\quad 
\b{F}_\Phi = -m y.
\end{equation}
Substituting the solutions into the Lagrangian in \er{181013.2038},
we obtain
\begin{equation}
\begin{split}
 {\cal L}_{\rm top., SUSY}
& = 
-\der^m y \der_m \b{y}
- \fr{1}{2} \der^m \phi \der_m \phi 
- \fr{1}{2} \der^m s \der_m s 
\\
&
\quad
- m^2 |y|^2 
-\fr{1}{2}m^2 s^2 
-\fr{1}{2}(m\phi+ c)^2.
\end{split} 
\label{181013.2039}
\end{equation}
The last term is the mass term for the pseudo-scalar field.
Note that the mass term for the scalar field $s$ is 
also generated due to SUSY.

\subsection{Scalar potential from higher derivative terms}
\label{sth}
In this subsection, we consider the generation of the potentials for the pseudo-scalar field from the topological coupling and 
the ghost/tachyon-free higher derivative terms of the 3-form gauge field, 
which is our new result in this paper.

In the previous sections \ref{btc} and \ref{stc}, 
we have seen that the quadratic potential (mass term) can be generated 
by the quadratic (canonical) kinetic term of the 3-form gauge field
in the bosonic and SUSY 3-form gauge theories, respectively.
Furthermore, we have reviewed in section \ref{bth} that 
 more general potentials can be generated by 
higher derivative terms of the 3-form gauge field.
Thus, we can consider the potential generation mechanism 
in SUSY field theories.

Since a K\"ahler potential and a superpotential can only generate
 at most quadratic terms of the field strength,
we should consider higher derivative terms of the 3-form 
gauge field.
We thus use the higher derivative term of the 3-form gauge field 
proposed in Ref.~\cite{Nitta:2018yzb}.
This higher derivative term 
gives us the terms of an arbitrary order of the field strength.
Note that some higher derivative terms of the 3-form gauge field 
 such as $\der^m F\der_m F$ may cause tachyons and/or ghosts,
although the higher derivative terms which we will use 
are free from such instabilities.

\subsubsection{General arguments}
Here, we consider a general expression of the scalar potential 
generated by the topological coupling and the higher derivative 
terms of the 3-form gauge field.
A Lagrangian with the topological coupling and 
higher derivative terms is given by%
\footnote{In this Lagrangian, 
we consider the simplest K\"ahler potential 
$\fr{1}{2}(\Phi+\b\Phi)^2 + Y\b{Y}$,
although this quardratic kinetic term  
can be exnteneded to a general form $K(\Phi+\b\Phi, Y, \b{Y})$
while preserving the shift symmetry of the chiral superfield $\Phi$.
The chiral superfield $\Phi$ can be dualized into 
a deformed real linear multiplet $L$ satisfying $L^\dg = L$ and
$-\fr{1}{4} \b{D}^2 L = mY$ 
(see e.g.~\cite{Dudas:2014pva,Kuzenko:2017oni}).
The authors thank S.~M.~Kuzenko for comments on the generalization
of the K\"ahler potential and the dual transformation.}
\begin{equation}
\begin{split}
 {\cal L}_{\rm top.,HD,SUSY}
& = 
-\df{1}{8} \int d^2\theta \b{D}^2 
\(\fr{1}{2}(\Phi+\b\Phi)^2 + Y\b{Y}\)
+ m \int d^2\theta \Phi Y
\\
&
\quad
-\df{1}{8\cdot 16} \int d^2\theta \b{D}^2 \Lambda
(D^\alpha Y)(D_\alpha Y)(\b{D}_{\d\alpha} \b{Y})
(\b{D}^{\d\alpha} \b{Y})
\\
&
\quad
+\fr{i}{8} 
\(
\int d^2 \theta \b{D}^2
-\int d^2\b\theta D^2
\) T_{I, \rm top.,HD}  X
+\hc 
\end{split}
\label{181014.0128}
\end{equation}
Here, $\Lambda$ is a real function of $D^2 Y$ and $\b{D}^2 \b{Y}$:
\begin{equation}
 \Lambda = \Lambda \( D^2 Y, \b{D}^2 \b{Y}\).
\end{equation}
The second line in \er{181014.0128} 
is the higher derivative term for the 
3-form gauge field.
The last line in \er{181014.0128} is 
the boundary term for the Lagrangian.
The superfield $T_{I, \rm top., HD}$ in the last line is 
the imaginary part of the following chiral superfield
\begin{equation}
\begin{split}
 T_{\rm top., HD}
&= 
 -\fr{1}{4} \b{D}^2 \b{Y} + m \Phi
\\
&
\quad
-\df{1}{4\cdot 16}\b{D}^2
\Bigg(
\pd{\Lambda}{Y} |D_\alpha Y|^4
+D^2 \(\pd{\Lambda}{D^2 Y} |D_\alpha Y|^4 \)
\\
&
\hph{\quad
-\df{1}{4}\b{D}^2\cdot \df{1}{16}
\Bigg(}
\quad
-2 D^\alpha 
(\Lambda (D_\alpha Y)(\b{D}_{\d\alpha} \b{Y}) (\b{D}^{\d\alpha}\b{Y}))
\Bigg).
\end{split}
\end{equation}
The superfield can be determined by requiring 
that the functional variation of the 3-form prepotential at 
the boundary should vanish~\cite{Nitta:2018yzb}.
The bosonic part of the
$\theta = \b\theta =0$ component 
of the chiral superfield $ T_{\rm top., HD}$ is 
\begin{equation}
\begin{split}
 T_{\rm top., HD}|
&
= 
\b{\cal F} +\df{1}{\sr{2}}m (s + i\phi) 
+ 
2\Lambda |{\cal F}|^2 \b{\cal F}
-2\Lambda \b{\cal F} \der^m y \der_m \b{y}
\\
&
\quad
-4 \(\pd{\Lambda}{D^2 Y} |\)
\(
|{\cal F}|^4 -2 \der^m y \der_m \b{y} |{\cal F}|^2
+(\der^m y \der_m y )(\der^n \b{y} \der_n \b{y} )
\).
\end{split}
\end{equation}
The Lagrangian in \er{181014.0128} is invariant under the 
shift of the chiral superfield $\Phi \to \Phi + i k$,
with a real constant  $k $. 
This is because the higher derivative term only depends on 
$Y$, and does not break this shift symmetry of $\Phi$.

In order to discuss the potential for the pseudo-scalar field, 
we show the bosonic part of the Lagrangian in \er{181014.0128}:
\begin{equation}
\begin{split}
 {\cal L}_{\rm top., HD, SUSY}
& = 
-\der^m y \der_m \b{y}
+{\cal F}\b{\cal F}
- \fr{1}{2} \der^m \phi \der_m \phi 
- \fr{1}{2} \der^m s \der_m s 
+ F_\Phi \b{F}_{\Phi}
\\
&
\quad
+ m (y F_\Phi + \b{y} \b{F}_\Phi 
+{\cal F} \varphi +\b{\cal F} \b\varphi
) 
\\
&
\quad
+\Lambda 
\Big(|{\cal F}|^4 - 2 |{\cal F}|^2 \der^n y \der_n \b{y} 
+ (\der^n y \der_n y)(\der^p \b{y} \der_p \b{y}) \Big)
\\
&
\quad
 - \df{\sr{2}}{3!}
\der_m  \(T_{I,\rm top.,HD} |
\epsilon^{mnpq} C_{npq}
\).
\end{split} 
\label{181014.0127}
\end{equation}
Here, we have used the 
complex auxiliary field ${\cal F}$, instead of $F$ and $H$,
since it will be convenient to see the potentials for the 
pseudo-scalar field $\phi$ and its superpartner $s$.

Now, we see the mechanism of the potential generation
in the higher derivative SUSY 3-form gauge theories.
To obtain the potential, we solve the EOM for 
the field strength and auxiliary fields 
as in section~\ref{stc}.
The solution to the EOM for the auxiliary field $F_\Phi$ 
is the same: $F_\Phi = -m \b{y}$.
While, the solutions to the EOM for $F$ and $H$
are deformed by the higher derivative terms as
\begin{equation}
\begin{split}
- \fr{1}{\sr{2}}ic =  
& 
\b{\cal F}
+ m \varphi
+
\pd{\Lambda}{{\cal F}}
(|{\cal F}|^4 -2 |{\cal F}|^2 (\der^m y \der_m \b{y})
+
(\der^m y \der_m y)(\der^n \b{y} \der_n \b{y} ))
\\
&
+2 \Lambda (|{\cal F}|^2\b{\cal F} - \b{\cal F} \der^m y\der_m \b{y}),
\end{split}
\label{181022.2138}
\end{equation}
where $c$ is determined by the boundary conditions.
Note that the EOM and thier solutions for $F$ and $H$ can be 
derived by the variation of the 3-form pretpotential $X$ in the
superspace~\cite{Nitta:2018yzb}.
The on-shell Lagrangian is therefore
\begin{equation}
\begin{split}
 {\cal L}_{\rm top., HD, SUSY}
& = 
-\der^m y \der_m \b{y}
+{\cal F}\b{\cal F}
- \fr{1}{2} \der^m \phi \der_m \phi 
- \fr{1}{2} \der^m s \der_m s 
- m^2 |y|^2
\\
&
\quad
+{\cal F} \(m \varphi + \fr{1}{\sr{2}} i c\)
+\b{\cal F} \(m \b\varphi - \fr{1}{\sr{2}} i c\)
\\
&
\quad
+\Lambda 
\Big(|{\cal F}|^4 - 2 |{\cal F}|^2 \der^n y \der_n \b{y} 
+ (\der^n y \der_n y)(\der^p \b{y} \der_p \b{y}) \Big),
\end{split} 
\label{181022.2139}
\end{equation}
where we have implicitly substituted the solution in \er{181022.2138}
into ${\cal F}$ in the on-shell Lagrangian.

We consider the relation between the higher derivative terms
of the 3-form gauge field and the generated potentials 
for the pseudo-scalar field.
The potential term can be simply seen by setting spacetime derivatives
on the following fields to zero:
$\der_m y = \der_m \b{y}= \der_m s = \der_m \phi=0$.
When the derivatives on the fields are set to zero, 
the Lagrangian in \er{181014.0127} becomes   
\begin{equation}
\begin{split}
 {\cal L}_{\rm top., HD, SUSY}
& = 
{\cal G}
 + F_\Phi \b{F}_{\Phi}
+ m (y F_\Phi + \b{y} \b{F}_\Phi 
+{\cal F} \varphi +\b{\cal F} \b\varphi
) 
\\
&
\quad
 - \df{\sr{2}}{3!}
\der_m  \(
\epsilon^{mnpq} C_{npq} 
{\rm Im}
\( \pd{{\cal G}}{{\cal F}} + m\varphi\)
\).
\end{split} 
\label{181024.0255} 
\end{equation}
Here, ${\cal G} = {\cal G}({\cal F}, \b{\cal F})$
is a real function of ${\cal F} $ and $\b{\cal F}$ defined by
\begin{equation}
{\cal G} = 
{\cal F}\b{\cal F}
+\Lambda |{\cal F}|^4,
\end{equation}
which can be understood as a SUSY  
extension of $G(F)$ in 
\er{180725.2141}. 
The solution to the auxiliary field ${\cal F}$ is simplified as
\begin{equation}
\begin{split}
-\fr{1}{\sr{2}} ic
&
 = 
\b{\cal F}
 +  m \varphi
+
\pd{\Lambda}{{\cal F}}
|{\cal F}|^4
+2 \Lambda |{\cal F}|^2\b{\cal F}.
 \end{split}
\label{181022.2152}
\end{equation}
This leads to the relation between ${\cal F}$ 
and the scalar field $\varphi$ as
\begin{equation}
-\fr{1}{\sr{2}} i c -m\varphi = 
\pd{{\cal G}}{{\cal F}}.
\label{181024.0435}
\end{equation}
Thus, the potential part of the on-shell Lagrangian is 
\begin{equation}
\begin{split}
 {\cal L}_{\rm top., HD, SUSY}
& = 
{\cal G}
-{\cal F} \pd{{\cal G}}{{\cal F}}
-\b{\cal F} \pd{{\cal G}}{\b{\cal F}}
-m^2 |y|^2.
\end{split} 
\label{181024.0417}  
\end{equation}
We thus obtain the relation between the on-shell potential of the 
scalar field ${\cal V}(\varphi, \b\varphi)$ and the 
function of the field strength ${\cal G}$ as
\begin{equation}
{\cal V} =  - {\cal G}
+{\cal F} \pd{{\cal G}}{{\cal F}}
+\b{\cal F} \pd{{\cal G}}{\b{\cal F}},
\label{181024.0426}
\end{equation}
which can be seen as a SUSY extension of the relation 
between the potential for the pseudo-scalar field $V(\phi)$
and the function of the field strength $G(F)$ in \er{181024.0427}.
Note that the relation between the auxiliary field ${\cal F}$ 
and the potential can be obtained from \er{181024.0435} and 
the derivatives 
of the both sides of \er{181024.0426} by $\varphi$:
\begin{equation}
 \pd{{\cal V}}{\varphi}
= -m {\cal F},
\label{190410.1603}
\end{equation}
which is a SUSY extension of \er{181024.0444}.
\subsubsection{Example 1: $\Lambda = \text{constant}$}
\begin{figure}[t]
 \begin{center}
  \includegraphics[width=18em]{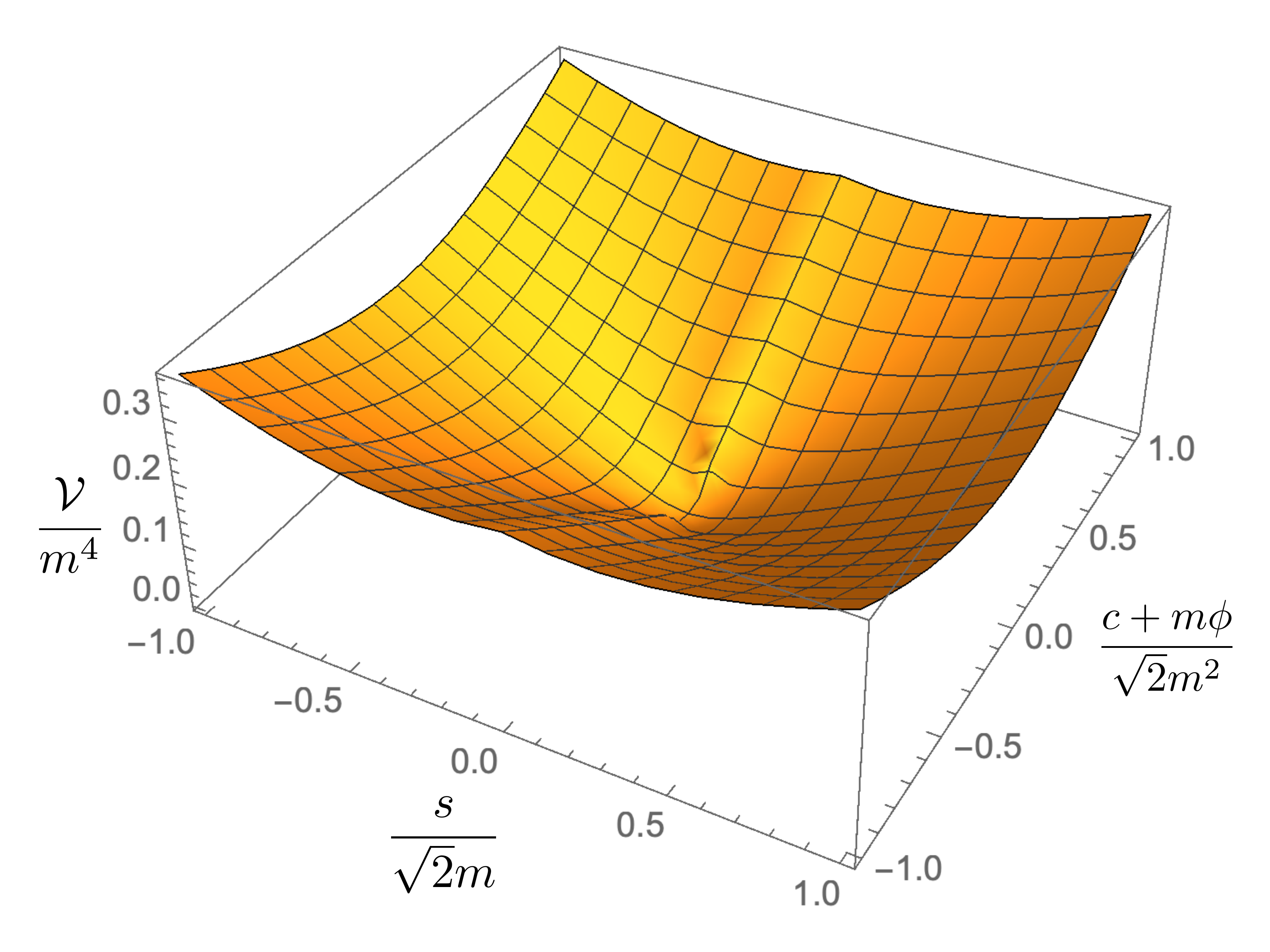}
  \includegraphics[width=18em]{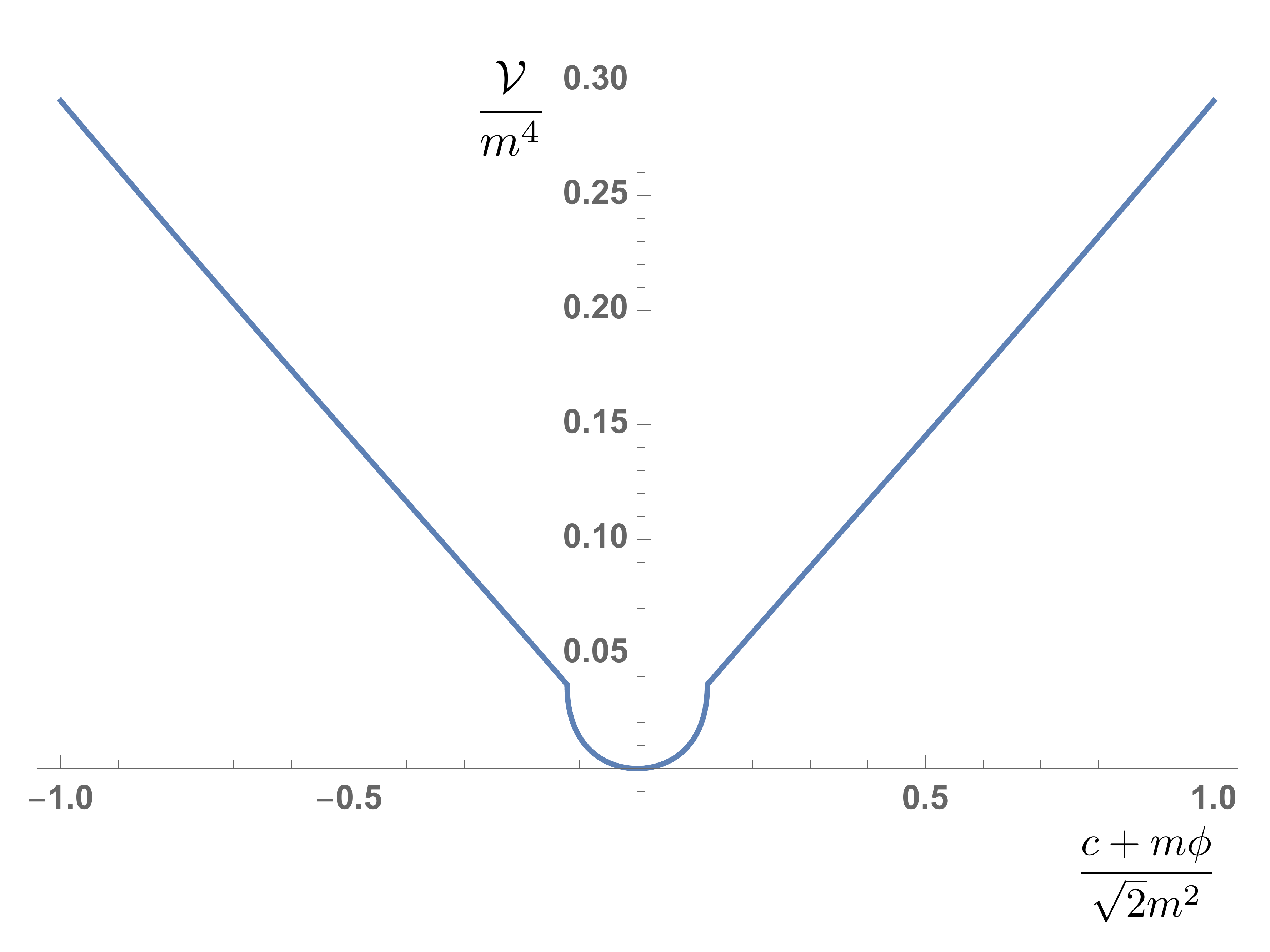}
 \end{center}
\caption{\footnotesize
On-shell potential for $k=0$ with $m^4\Lambda_0 = 10$. The right figure corresponds to $s=0$ in the left figure.}
\label{k0}
\end{figure}

\begin{figure}[t]
 \begin{center}
  \includegraphics[width=18em]{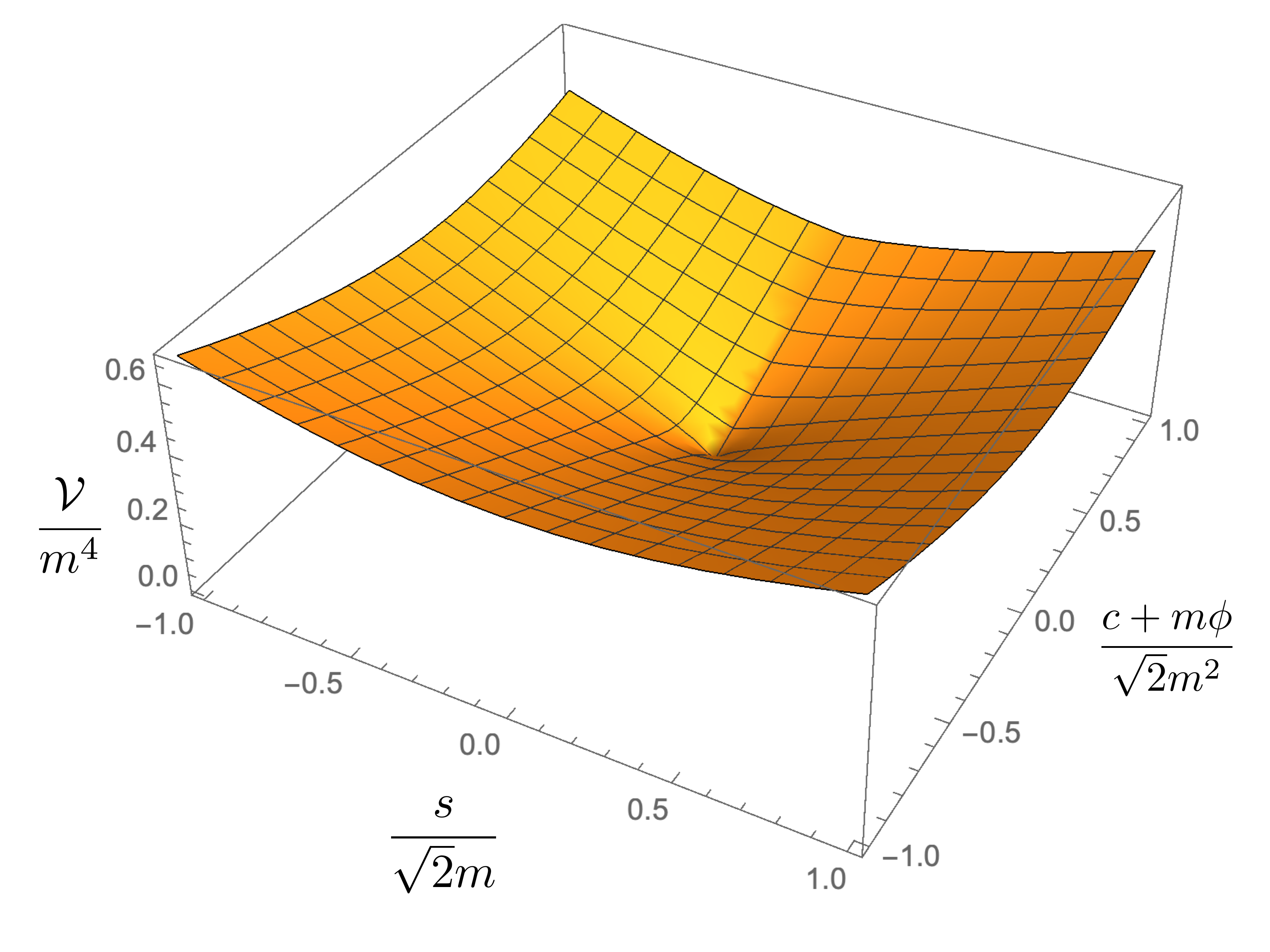}
  \includegraphics[width=18em]{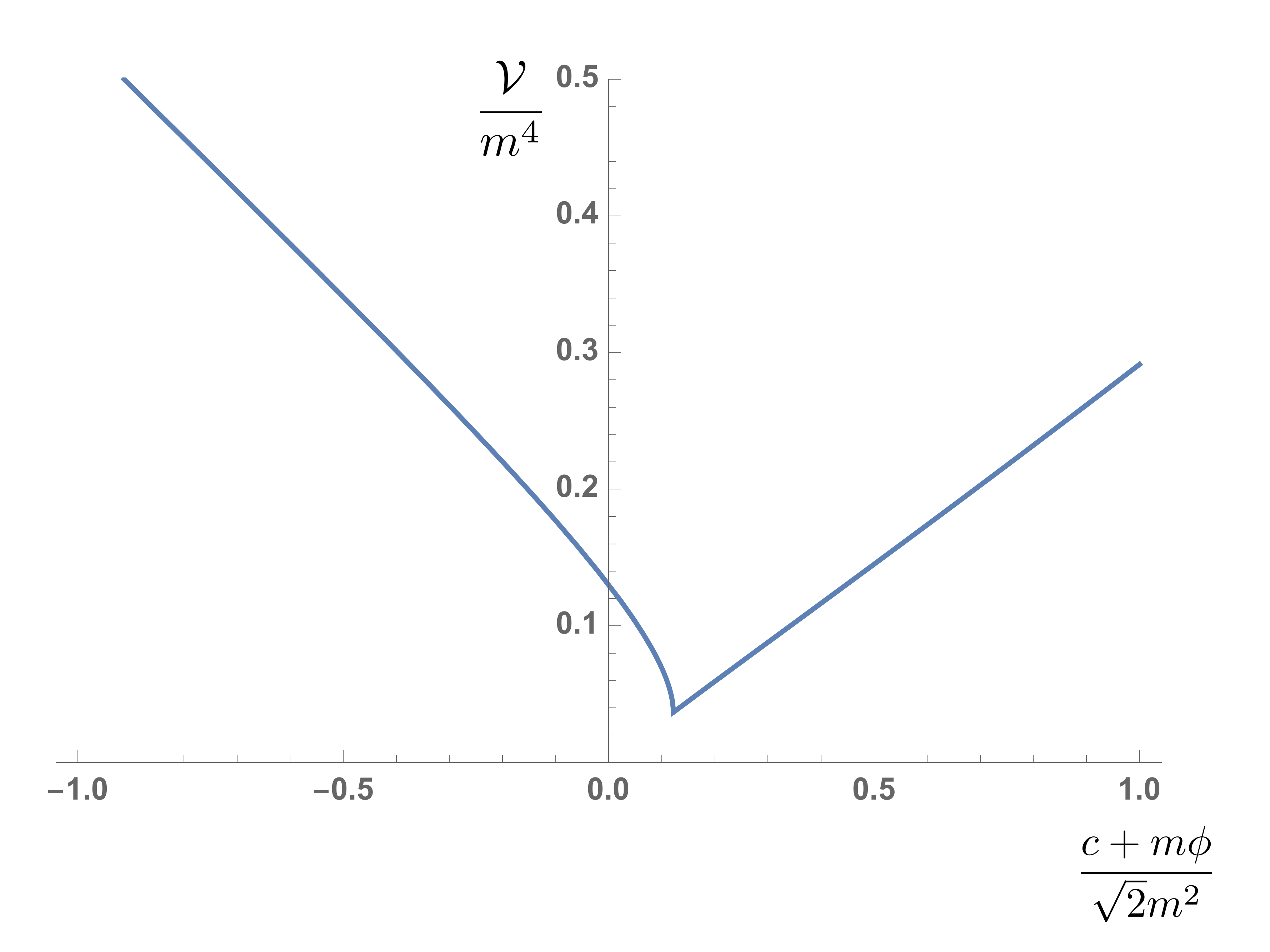}
 \end{center}
\caption{\footnotesize
On-shell potential for $k=1$ with $m^4\Lambda_0 = 10$. 
The right figure corresponds to $s=0$ in the left figure.}
\label{k1}
\end{figure}

As a simple example, we consider the case where $\Lambda$ is 
a constant: $\Lambda = \Lambda_0$, which was described in 
Ref.~\cite{Dudas:2014pva} and we will 
explicitly solve the EOM.
In this choice of $\Lambda$, ${\cal G}$ is given by
\begin{equation}
{\cal G} = |{\cal F}|^2 + \Lambda_0 |{\cal F}|^4,
\end{equation}
while the on-shell potential ${\cal V}$ is 
\begin{equation}
 {\cal V} = |{\cal F}|^2 + 3\Lambda_0 |{\cal F}|^4.
\label{181024.2129}
\end{equation}
In order to obtain the potential, 
we express ${\cal F}$ in terms of $\varphi$ by solving the EOM.
The EOM for ${\cal F}$ is 
\begin{equation}
 -\fr{1}{\sr{2}} ic -  m\varphi  
= \b{\cal F} +2\Lambda_0 |{\cal F}|^2 \b{\cal F}.
\label{181024.1648}
\end{equation}
By using \er{181024.1648}, 
the auxiliary field ${\cal F} $ can be related to 
the Hermitian conjugate $\b{\cal F}$ as
\begin{equation}
{\cal F} = 
- \fr{ \fr{1}{\sr{2}} ic + m\varphi + \b{\cal F}}
{2\Lambda_0  \b{\cal F}^2}. 
\end{equation}
Substituting the Hermitian conjugate of the relation,
\begin{equation}
\b{\cal F} = 
- \fr{ -\fr{1}{\sr{2}} ic + m\b\varphi + {\cal F}}
{2\Lambda_0  {\cal F}^2},
\end{equation}
into \er{181024.1648}, we obtain
\begin{equation}
 2\Lambda_0  \(\fr{1}{\sr{2}}ic + m\varphi\) {\cal F}^3
+
 \(-\fr{1}{\sr{2}}ic + m\b\varphi\) {\cal F}
+
 \(-\fr{1}{\sr{2}}ic + m\b\varphi\)^2 =0.
\label{190410.1611}
\end{equation}
This can be solved by using the Cardano's formula~\cite{Sasaki:2012ka}
\begin{equation}
 {\cal F} 
=  
e^{-i\eta}
 \omega^k
\sr[3]{-\fr{q}{2} + \sr{\(\fr{q}{2}\)^2 + \(\fr{p}{3}\)^3}}
-
e^{-i\eta}\omega^{3-k}
\sr[3]{\fr{q}{2} + \sr{\(\fr{q}{2}\)^2 + \(\fr{p}{3}\)^3}}
,
\label{181024.2128}
\end{equation}
where 
\begin{equation}
\begin{split}
&
m\varphi +\fr{ic}{\sr{2}} = re^{i\eta},
\quad
 p = \fr{1}{ 2\Lambda_0 },
\quad
q=\fr{r}{ 2\Lambda_0 }, 
\quad
\omega^3= 1,
\quad
k=0,1,2. 
\end{split}
\end{equation}
Here, $r>0$ and $\eta $ are real parameters.
The on-shell potential for the scalar field ${\cal V}$ is
the one obtained by substituting the solution in \er{181024.2128} into
the potential in \er{181024.2129}.
Examples of the on-shell potentials are found in Figure~\ref{k0} and \ref{k1}
for $k=0,1$, respectively (the potential for $k=2$ can be obtained by 
$c + m\phi \to -(c+m\phi)$ of the potential for $k=1$).

The higher derivative terms generally deform the vacuum structure.
In particular, SUSY can be spontaneously broken in the vacuum
by the effect of the higher derivative terms~\cite{Antoniadis:2007xc}.
Here, we discuss the vacuum structure in this model.
In the following discussion, 
we assume $\Lambda_0 > 0$ for the stability of the system.

As in ordinary chiral superfield cases,
whether SUSY is preserved or broken can be examined by 
the condition whether
${\cal F} =0$ is consistent with the EOM of ${\cal F}$ or not.
To examine it, let us assume ${\cal F} =0$.
By the EOM for ${\cal F}$ in \er{181024.1648}, 
the assumption ${\cal F} = 0$ leads to the 
condition for $\varphi$:
\begin{equation}
 \varphi = -\fr{ic}{\sr{2}m}.
\label{190412.0915}
\end{equation}
Substituting this condition into the EOM in \er{181024.2128},
one can show that the branch $k=0$ 
(called the canonical branch) is only consistent with 
the condition in \er{190412.0915}, 
while the branches $k = 1,2$ 
(called the non-canonical branches) are not.
This can be seen as follows.
We express $\varphi$ around the condition in \er{190412.0915}
as $\varphi = -\fr{ic}{\sr{2}m} + \epsilon e^{i\eta_0}$, 
where $\epsilon$ is a real small parameter 
$\epsilon/m \ll 1$, and $\eta_0 $ is an angle.
The vacuum corresponds to $\epsilon =0$.
The solution of the EOM in \er{181024.2128} can be now expressed 
as
\begin{equation}
 {\cal F} = \omega^k e^{-i\eta_0}\df{1}{\sr{6\Lambda_0}} 
-\omega^{3-k} e^{-i\eta_0}\df{1}{\sr{6\Lambda_0}} +{\cal O}(\epsilon),
\end{equation}
and the branch $k=0$ only 
gives us the consistent solution ${\cal F} =0$.
Therefore, the SUSY is preserved in the branch $k=0$,
while SUSY is spontaneously broken in the branches $k = 1,2$.
Note that the values of ${\cal F}$ in the branches $k = 1,2$
are ${\cal F} = e^{-i\eta_0} \fr{i}{\sr{2\Lambda_0}}, 
-e^{-i\eta_0} \fr{i}{\sr{2\Lambda_0}}$, respectively, and 
 the phase of ${\cal F}$ in the vacuum depends on 
 the direction of a limit approaching to 
 the vacuum in  Eq.~(\ref{190412.0915}). 

\subsubsection{Example 2: Cosine-type potential}
Next, we consider a cosine potential in SUSY 3-form gauge theories.
In order to generate the cosine potential, 
we should choose a suitable real function $\Lambda$ 
so that the function ${\cal G}$ has the same structure as 
the function in \er{181025.0814}.
We choose it as follows:
\begin{equation}
 \Lambda = 
\df{1}{|-\tfrac{1}{4} D^2 Y|^4} 
\(M^2 {\hat{F}} \arcsin\(\df{\hat{F}}{M^2}\)
+ M^4\sr{1- \df{\hat{F}^2}{M^4}} 
-\fr{1}{2} \hat{F}^2
\).
\end{equation}
Here, $\hat{F}$ is defined by
\begin{equation}
 \hat{F}:= -\df{\sr{2}i}{8}(D^2 Y-\b{D}^2\b{Y}).
\end{equation}
The $\theta = \b\theta =0$ component of $\hat{F}$ is identified as 
the field strength $F$:
\begin{equation}
 \hat{F}|= -\df{\sr{2}i}{8}(D^2 Y-\b{D}^2\b{Y})| = F. 
\end{equation}
In this choice, the function ${\cal G}$ is given by
\begin{equation}
 {\cal G} = 
M^2 F \arcsin\(\df{F}{M^2}\)
+ M^4\sr{1- \df{F^2}{M^4}} + \fr{1}{2}H^2,
\end{equation} 
which is the same as the function in \er{181025.0814}.
The on-shell potential in \er{181024.0426} is therefore
\begin{equation}
 {\cal V}
= - M^4 \sr{1- \fr{F^2}{M^4}} + \fr{1}{2} H^2.
\label{181025.0947} 
\end{equation}
Note that we can choose the desired function ${\cal G}$ 
thanks to the denominator $|-\fr{1}{4} D^2 Y|^4$ in $\Lambda$.

Now, we express the on-shell potential ${\cal V}$ in terms of $\varphi$
by solving the EOM for ${\cal F}$.
The EOM for ${\cal F}$ is 
\begin{equation}
-\fr{1}{\sr{2}} i c -m\varphi = 
\pd{{\cal G}}{{\cal F}} = 
\pd{H}{\cal F} \pd{\cal G}{H}
+
\pd{F}{\cal F} \pd{\cal G}{F}
=
\fr{1}{\sr{2}} H
-\df{i}{\sr{2}}M^2 \arcsin \(\fr{F}{M^2}\).
\label{181024.0435-2}
\end{equation}
The real part of the equation gives us
\begin{equation}
 H  = -ms,
\end{equation} 
while the imaginary part gives us
\begin{equation}
 F = -M^2 \sin\(\fr{m\phi + c}{M^2}\).
\end{equation}
Substituting the solutions into \er{181025.0947}, 
we obtain the Lagrangian with the following potential 
for $s$ and $\phi$:
\begin{equation}
  {\cal V} = - M^4 \cos \(\df{m\phi +c }{M^2}\) + \fr{1}{2} m^2 s^2.
\end{equation}
We thus have obtained the cosine potential for the pseudo-scalar field 
in addition to the mass term for $s$.

\section{Summary}
\label{sum}

In this paper, we have discussed a 
topological coupling of a 3-form gauge field with 
a pseudo-scalar field in SUSY higher derivative 3-form gauge theories.
We have used a ghost/tachyon-free higher derivative Lagrangian of 
the 3-form gauge field.
We have shown that the potential for the pseudo-scalar field is generated
by substituting the solution of the EOM for the 3-form gauge field.
We have presented two examples of the higher derivative Lagrangians
and the corresponding potentials for the pseudo-scalar field.
One is the quartic order of the field strength, which was
described in Ref.~\cite{Dudas:2014pva}.
Another example is a SUSY extension of the bosonic model proposed in Ref.~\cite{Dvali:2005an}, which generates a cosine potential for the pseudo-scalar field.

There can be some applications of the above Lagrangians.
For example,
we can apply them to SUGRA inflationary models.
In this application, inflaton can be identified as the pseudo-scalar 
field. Since the pseudo-scalar field is shift-symmetric in the Lagrangian discussed in this paper, it can be a candidate for the origin for flatness of the inflaton potential.
Furthermore, the higher derivative terms of the 3-form gauge field may play an important role to flatten the inflaton potential in SUGRA inflationary models
as in the bosonic models in Ref.~\cite{DAmico:2017cda}.

In order to apply the mechanism of the 
generation of the potentials to the inflationary models, 
we should embed the topological coupling and the higher 
derivative terms into SUGRA.
It will be convenient 
to use conformal SUGRA~\cite{Cremmer:1982en,Kugo:1982cu,Kugo:1982mr,Kugo:1983mv,Butter:2009cp,Kugo:2016zzf,Kugo:2016lum},
since we can obtain the canonically normalized Einstein--Hilbert term 
by the superconformal gauge-fixing 
and avoid a tedious super-Weyl rescaling~\cite{Kugo:1982mr}.
We will address the issue elsewhere.

In this paper, we have considered the potentials
generated by the topological coupling and the higher derivative 
terms, but not considered the kinetic terms.
In SUSY higher derivative theories of chiral superfields, 
it is known that kinetic terms in on-shell Lagrangians
can also be deformed by higher derivative terms.
If $\Lambda $ is a constant as in the example~1 in Sec.~\ref{sth}, 
there are solutions to the auxiliary fields (so-called 
non-canonical branches) 
where the on-shell canonical kinetic term for the scalar field $y$
vanishes ~\cite{Khoury:2010gb,Koehn:2012te,Sasaki:2012ka,Farakos:2012qu,Adam:2013awa,Nitta:2014pwa}
in addition to the conventional solution ${\cal F}=0$
 (the canonical branch) giving rise to the usual kinetic term.
It would be interesting to consider how the solutions are 
deformed in the presence of the topological coupling and 
the higher derivative terms with an arbitrary order of 
the field strength, which we have used in this paper.

\subsection*{Acknowledgements}
This work is supported by the Ministry of Education, 
Culture, Sports, Science (MEXT)-Supported
Program for the Strategic Research Foundation  at
Private Universities ``Topological Science'' (Grant No.~S1511006). 
M.~N.~is also supported in part by JSPS Grant-in-Aid
for Scientific Research (KAKENHI Grant No.~16H03984 and No.~18H01217),
and also by MEXT KAKENHI Grant-in-Aid for Scientific 
Research on Innovative Areas ``Topological Materials Science'' 
No.~15H05855.

\providecommand{\href}[2]{#2}
\begingroup\begin{thebibliography}{99}

\bibitem{Aurilia:1977jz}
A.~Aurilia and F.~Legovini, ``{{Extended Systems and Generalized London
  Equations}},''\href{https://doi.org/10.1016/0370-2693(77)90376-8}{ {Phys.
  Lett.} {\bf 67B} (1977) 299--302}.

\bibitem{Aurilia:1978dw}
A.~Aurilia, ``{{The Problem of Confinement: From Two-dimensions to
  Four-dimensions}},''\href{https://doi.org/10.1016/0370-2693(79)90524-0}{
  {Phys. Lett.} {\bf 81B} (1979) 203--206}.

\bibitem{Luscher:1978rn}
M.~Luscher, ``{{The Secret Long Range Force in Quantum Field Theories With
  Instantons}},''\href{https://doi.org/10.1016/0370-2693(78)90487-2}{ {Phys.
  Lett.} {\bf 78B} (1978) 465--467}.

\bibitem{Aurilia:1980jz}
A.~Aurilia, Y.~Takahashi, and P.~K. Townsend, ``{{The U(1) Problem and the
  Higgs Mechanism in Two-dimensions and
  Four-dimensions}},''\href{https://doi.org/10.1016/0370-2693(80)90484-0}{
  {Phys. Lett.} {\bf 95B} (1980) 265--268}.

\bibitem{Hata:1980hn}
H.~Hata, T.~Kugo, and N.~Ohta, ``{{Skew Symmetric Tensor Gauge Field Theory
  Dynamically Realized in {QCD} U(1)
  Channel}},''\href{https://doi.org/10.1016/0550-3213(81)90170-X}{ {Nucl.
  Phys.} {\bf B178} (1981) 527--544}.

\bibitem{Dvali:2005an}
G.~Dvali, ``{{Three-form gauging of axion symmetries and gravity}},''
  \href{https://arxiv.org/abs/hep-th/0507215}{arXiv:hep-th/0507215 [hep-th]}.

\bibitem{Dvali:2005zk}
G.~Dvali, ``{{A Vacuum accumulation solution to the strong CP
  problem}},''\href{https://doi.org/10.1103/PhysRevD.74.025019}{ {Phys. Rev.}
  {\bf D74} (2006) 025019},
  [\href{https://arxiv.org/abs/hep-th/0510053}{arXiv:hep-th/0510053 [hep-th]}].

\bibitem{Aurilia:1980xj}
A.~Aurilia, H.~Nicolai, and P.~K. Townsend, ``{{Hidden Constants: The Theta
  Parameter of QCD and the Cosmological Constant of N=8
  Supergravity}},''\href{https://doi.org/10.1016/0550-3213(80)90466-6}{ {Nucl.
  Phys.} {\bf B176} (1980) 509--522}.

\bibitem{Hawking:1984hk}
S.~W. Hawking, ``{{The Cosmological Constant Is Probably
  Zero}},''\href{https://doi.org/10.1016/0370-2693(84)91370-4}{ {Phys. Lett.}
  {\bf 134B} (1984) 403}.

\bibitem{Brown:1987dd}
J.~D. Brown and C.~Teitelboim, ``{{Dynamical Neutralization of the Cosmological
  Constant}},''\href{https://doi.org/10.1016/0370-2693(87)91190-7}{ {Phys.
  Lett.} {\bf B195} (1987) 177--182}.

\bibitem{Brown:1988kg}
J.~D. Brown and C.~Teitelboim, ``{{Neutralization of the Cosmological Constant
  by Membrane Creation}},''\href{https://doi.org/10.1016/0550-3213(88)90559-7}{
  {Nucl. Phys.} {\bf B297} (1988) 787--836}.

\bibitem{Duff:1989ah}
M.~J. Duff, ``{{The Cosmological Constant Is Possibly Zero, but the Proof Is
  Probably Wrong}},''\href{https://doi.org/10.1016/0370-2693(89)90284-0}{
  {Phys. Lett.} {\bf B226} (1989) 36}. [Conf. Proc.C8903131,403(1989)].

\bibitem{Duncan:1989ug}
M.~J. Duncan and L.~G. Jensen, ``{{Four Forms and the Vanishing of the
  Cosmological
  Constant}},''\href{https://doi.org/10.1016/0550-3213(90)90344-D}{ {Nucl.
  Phys.} {\bf B336} (1990) 100--114}.

\bibitem{Kaloper:2008qs}
N.~Kaloper and L.~Sorbo, ``{{Where in the String Landscape is
  Quintessence}},''\href{https://doi.org/10.1103/PhysRevD.79.043528}{ {Phys.
  Rev.} {\bf D79} (2009) 043528},
  [\href{https://arxiv.org/abs/0810.5346}{arXiv:0810.5346 [hep-th]}].

\bibitem{DAmico:2018mnx}
G.~D'Amico, N.~Kaloper, and A.~Lawrence, ``{{Strongly Coupled Quintessence}},''
  \href{https://arxiv.org/abs/1809.05109}{arXiv:1809.05109 [hep-th]}.

\bibitem{Kaloper:2008fb}
N.~Kaloper and L.~Sorbo, ``{{A Natural Framework for Chaotic
  Inflation}},''\href{https://doi.org/10.1103/PhysRevLett.102.121301}{ {Phys.
  Rev. Lett.} {\bf 102} (2009) 121301},
  [\href{https://arxiv.org/abs/0811.1989}{arXiv:0811.1989 [hep-th]}].

\bibitem{Kaloper:2011jz}
N.~Kaloper, A.~Lawrence, and L.~Sorbo, ``{{An Ignoble Approach to Large Field
  Inflation}},''\href{https://doi.org/10.1088/1475-7516/2011/03/023}{ {JCAP}
  {\bf 1103} (2011) 023},
  [\href{https://arxiv.org/abs/1101.0026}{arXiv:1101.0026 [hep-th]}].

\bibitem{Marchesano:2014mla}
F.~Marchesano, G.~Shiu, and A.~M. Uranga, ``{{F-term Axion Monodromy
  Inflation}},''\href{https://doi.org/10.1007/JHEP09(2014)184}{ {JHEP} {\bf 09}
  (2014) 184}, [\href{https://arxiv.org/abs/1404.3040}{arXiv:1404.3040
  [hep-th]}].

\bibitem{Kaloper:2014zba}
N.~Kaloper and A.~Lawrence, ``{{Natural chaotic inflation and ultraviolet
  sensitivity}},''\href{https://doi.org/10.1103/PhysRevD.90.023506}{ {Phys.
  Rev.} {\bf D90} (2014) no.~2, 023506},
  [\href{https://arxiv.org/abs/1404.2912}{arXiv:1404.2912 [hep-th]}].

\bibitem{Kaloper:2016fbr}
N.~Kaloper and A.~Lawrence, ``{{London equation for monodromy
  inflation}},''\href{https://doi.org/10.1103/PhysRevD.95.063526}{ {Phys. Rev.}
  {\bf D95} (2017) no.~6, 063526},
  [\href{https://arxiv.org/abs/1607.06105}{arXiv:1607.06105 [hep-th]}].

\bibitem{DAmico:2017cda}
G.~D'Amico, N.~Kaloper, and A.~Lawrence, ``{{Monodromy Inflation in the Strong
  Coupling Regime of the Effective Field
  Theory}},''\href{https://doi.org/10.1103/PhysRevLett.121.091301}{ {Phys. Rev.
  Lett.} {\bf 121} (2018) no.~9, 091301},
  [\href{https://arxiv.org/abs/1709.07014}{arXiv:1709.07014 [hep-th]}].

\bibitem{Ansoldi:1995by}
S.~Ansoldi, A.~Aurilia, and E.~Spallucci, ``{{Membrane vacuum as a type II
  superconductor}},''\href{https://doi.org/10.1142/S0217979296000775}{ {Int. J.
  Mod. Phys.} {\bf B10} (1996) 1695--1705},
  [\href{https://arxiv.org/abs/hep-th/9511096}{arXiv:hep-th/9511096 [hep-th]}].

\bibitem{Hasebe:2014nia}
K.~Hasebe, ``{{Higher Dimensional Quantum Hall Effect as A-Class Topological
  Insulator}},''\href{https://doi.org/10.1016/j.nuclphysb.2014.07.011}{ {Nucl.
  Phys.} {\bf B886} (2014) 952--1002},
  [\href{https://arxiv.org/abs/1403.5066}{arXiv:1403.5066 [hep-th]}].

\bibitem{Gates:1980ay}
S.~J. Gates, Jr., ``{{Super p-form gauge
  superfields}},''\href{https://doi.org/10.1016/0550-3213(81)90225-X}{ {Nucl.
  Phys.} {\bf B184} (1981) 381--390}.

\bibitem{Gates:1980az}
S.~J. Gates, Jr. and W.~Siegel, ``{{Variant superfield
  representations}},''\href{https://doi.org/10.1016/0550-3213(81)90281-9}{
  {Nucl. Phys.} {\bf B187} (1981) 389--396}.

\bibitem{Buchbinder:1988tj}
I.~L. Buchbinder and S.~M. Kuzenko, ``{{Quantization of the classically
  equivalent theories in the superspace of simple supergravity and quantum
  equivalence}},''\href{https://doi.org/10.1016/0550-3213(88)90047-8}{ {Nucl.
  Phys.} {\bf B308} (1988) 162--190}.

\bibitem{Binetruy:1996xw}
P.~Binetruy, F.~Pillon, G.~Girardi, and R.~Grimm, ``{{The Three form multiplet
  in supergravity}},''\href{https://doi.org/10.1016/0550-3213(96)00370-7}{
  {Nucl. Phys.} {\bf B477} (1996) 175--202},
  [\href{https://arxiv.org/abs/hep-th/9603181}{arXiv:hep-th/9603181 [hep-th]}].

\bibitem{Gates:1983nr}
S.~J. Gates, M.~T. Grisaru, M.~Rocek, and W.~Siegel, ``{{Superspace Or One
  Thousand and One Lessons in Supersymmetry}},''{ {Front. Phys.} {\bf 58}
  (1983) 1--548},
  [\href{https://arxiv.org/abs/hep-th/0108200}{arXiv:hep-th/0108200 [hep-th]}].

\bibitem{Buchbinder:1998qv}
I.~L. Buchbinder and S.~M. Kuzenko, ``{{Ideas and methods of supersymmetry and
  supergravity: Or a walk through superspace}},''{ {Bristol, UK: IOP (1998) 656
  p} (1998) }.

\bibitem{Binetruy:2000zx}
P.~Binetruy, G.~Girardi, and R.~Grimm, ``{{Supergravity couplings: A Geometric
  formulation}},''\href{https://doi.org/10.1016/S0370-1573(00)00085-5}{ {Phys.
  Rept.} {\bf 343} (2001) 255--462},
  [\href{https://arxiv.org/abs/hep-th/0005225}{arXiv:hep-th/0005225 [hep-th]}].

\bibitem{Binetruy:1995hq}
P.~Binetruy, M.~K. Gaillard, and T.~R. Taylor, ``{{Dynamical supersymmetric
  breaking and the linear
  multiplet}},''\href{https://doi.org/10.1016/0550-3213(95)00494-D}{ {Nucl.
  Phys.} {\bf B455} (1995) 97--108},
  [\href{https://arxiv.org/abs/hep-th/9504143}{arXiv:hep-th/9504143 [hep-th]}].

\bibitem{Binetruy:1995ta}
P.~Binetruy and M.~K. Gaillard, ``{{S duality constraints on effective
  potentials for gaugino
  condensation}},''\href{https://doi.org/10.1016/0370-2693(95)01242-7}{ {Phys.
  Lett.} {\bf B365} (1996) 87--97},
  [\href{https://arxiv.org/abs/hep-th/9506207}{arXiv:hep-th/9506207 [hep-th]}].

\bibitem{Dvali:1999pk}
G.~R. Dvali, G.~Gabadadze, and Z.~Kakushadze, ``{{BPS domain walls in large N
  supersymmetric QCD}},''\href{https://doi.org/10.1016/S0550-3213(99)00562-3}{
  {Nucl. Phys.} {\bf B562} (1999) 158--180},
  [\href{https://arxiv.org/abs/hep-th/9901032}{arXiv:hep-th/9901032 [hep-th]}].

\bibitem{Ovrut:1997ur}
B.~A. Ovrut and D.~Waldram, ``{{Membranes and three form
  supergravity}},''\href{https://doi.org/10.1016/S0550-3213(97)00510-5}{ {Nucl.
  Phys.} {\bf B506} (1997) 236--266},
  [\href{https://arxiv.org/abs/hep-th/9704045}{arXiv:hep-th/9704045 [hep-th]}].

\bibitem{Kuzenko:2017vil}
S.~M. Kuzenko and G.~Tartaglino-Mazzucchelli, ``{{Complex three-form
  supergravity and
  membranes}},''\href{https://doi.org/10.1007/JHEP12(2017)005}{ {JHEP} {\bf 12}
  (2017) 005}, [\href{https://arxiv.org/abs/1710.00535}{arXiv:1710.00535
  [hep-th]}].

\bibitem{Bandos:2018gjp}
I.~Bandos, F.~Farakos, S.~Lanza, L.~Martucci, and D.~Sorokin, ``{{Three-forms,
  dualities and membranes in four-dimensional
  supergravity}},''\href{https://doi.org/10.1007/JHEP07(2018)028}{ {JHEP} {\bf
  07} (2018) 028}, [\href{https://arxiv.org/abs/1803.01405}{arXiv:1803.01405
  [hep-th]}].

\bibitem{Kuzenko:2005wh}
S.~M. Kuzenko and S.~A. McCarthy, ``{{On the component structure of N=1
  supersymmetric nonlinear
  electrodynamics}},''\href{https://doi.org/10.1088/1126-6708/2005/05/012}{
  {JHEP} {\bf 05} (2005) 012},
  [\href{https://arxiv.org/abs/hep-th/0501172}{arXiv:hep-th/0501172 [hep-th]}].

\bibitem{Farakos:2016hly}
F.~Farakos, A.~Kehagias, D.~Racco, and A.~Riotto, ``{{Scanning of the
  Supersymmetry Breaking Scale and the Gravitino Mass in
  Supergravity}},''\href{https://doi.org/10.1007/JHEP06(2016)120}{ {JHEP} {\bf
  06} (2016) 120}, [\href{https://arxiv.org/abs/1605.07631}{arXiv:1605.07631
  [hep-th]}].

\bibitem{Farakos:2017jme}
F.~Farakos, S.~Lanza, L.~Martucci, and D.~Sorokin, ``{{Three-forms in
  Supergravity and Flux
  Compactifications}},''\href{https://doi.org/10.1140/epjc/s10052-017-5185-y}{
  {Eur. Phys. J.} {\bf C77} (2017) no.~9, 602},
  [\href{https://arxiv.org/abs/1706.09422}{arXiv:1706.09422 [hep-th]}].

\bibitem{Groh:2012tf}
K.~Groh, J.~Louis, and J.~Sommerfeld, ``{{Duality and Couplings of
  3-Form-Multiplets in N=1
  Supersymmetry}},''\href{https://doi.org/10.1007/JHEP05(2013)001}{ {JHEP} {\bf
  05} (2013) 001}, [\href{https://arxiv.org/abs/1212.4639}{arXiv:1212.4639
  [hep-th]}].

\bibitem{Hartong:2009az}
J.~Hartong, M.~Hubscher, and T.~Ortin, ``{{The Supersymmetric tensor hierarchy
  of N=1,d=4
  supergravity}},''\href{https://doi.org/10.1088/1126-6708/2009/06/090}{ {JHEP}
  {\bf 06} (2009) 090}, [\href{https://arxiv.org/abs/0903.0509}{arXiv:0903.0509
  [hep-th]}].

\bibitem{Becker:2016xgv}
K.~Becker, M.~Becker, W.~D. Linch, and D.~Robbins, ``{{Abelian tensor hierarchy
  in 4D, N = 1 superspace}},''\href{https://doi.org/10.1007/JHEP03(2016)052}{
  {JHEP} {\bf 03} (2016) 052},
  [\href{https://arxiv.org/abs/1601.03066}{arXiv:1601.03066 [hep-th]}].

\bibitem{Aoki:2016rfz}
S.~Aoki, T.~Higaki, Y.~Yamada, and R.~Yokokura, ``{{Abelian tensor hierarchy in
  4D ${\cal N}=1$ conformal
  supergravity}},''\href{https://doi.org/10.1007/JHEP09(2016)148}{ {JHEP} {\bf
  09} (2016) 148}, [\href{https://arxiv.org/abs/1606.04448}{arXiv:1606.04448
  [hep-th]}].

\bibitem{Dudas:2014pva}
E.~Dudas, ``{{Three-form multiplet and
  Inflation}},''\href{https://doi.org/10.1007/JHEP12(2014)014}{ {JHEP} {\bf 12}
  (2014) 014}, [\href{https://arxiv.org/abs/1407.5688}{arXiv:1407.5688
  [hep-th]}].

\bibitem{Yokokura:2016xcf}
R.~Yokokura, ``{{Abelian tensor hierarchy and Chern-Simons actions in 4D
  $\mathcal N=1$ conformal
  supergravity}},''\href{https://doi.org/10.1007/JHEP12(2016)092}{ {JHEP} {\bf
  12} (2016) 092}, [\href{https://arxiv.org/abs/1609.01111}{arXiv:1609.01111
  [hep-th]}].

\bibitem{Cribiori:2018jjh}
N.~Cribiori and S.~Lanza, ``{{On the dynamical origin of parameters in
  $\mathcal {N}=2$
  supersymmetry}},''\href{https://doi.org/10.1140/epjc/s10052-019-6545-6}{
  {Eur. Phys. J.} {\bf C79} (2019) no.~1, 32},
  [\href{https://arxiv.org/abs/1810.11425}{arXiv:1810.11425 [hep-th]}].

\bibitem{Buchbinder:2017vnb}
E.~I. Buchbinder and S.~M. Kuzenko, ``{{Three-form multiplet and supersymmetry
  breaking}},''\href{https://doi.org/10.1007/JHEP09(2017)089}{ {JHEP} {\bf 09}
  (2017) 089}, [\href{https://arxiv.org/abs/1705.07700}{arXiv:1705.07700
  [hep-th]}].

\bibitem{Kuzenko:2017oni}
S.~M. Kuzenko, ``{{Nilpotent ${\cal N}=1$ tensor
  multiplet}},''\href{https://doi.org/10.1007/JHEP04(2018)131}{ {JHEP} {\bf 04}
  (2018) 131}, [\href{https://arxiv.org/abs/1712.09258}{arXiv:1712.09258
  [hep-th]}].

\bibitem{Yamada:2018nsk}
Y.~Yamada, ``{{U(1) symmetric
  $\alpha$-attractors}},''\href{https://doi.org/10.1007/JHEP04(2018)006}{
  {JHEP} {\bf 04} (2018) 006},
  [\href{https://arxiv.org/abs/1802.04848}{arXiv:1802.04848 [hep-th]}].

\bibitem{Curtright:1980yj}
T.~L. Curtright and P.~G.~O. Freund, ``{{Massive dual
  fields}},''\href{https://doi.org/10.1016/0550-3213(80)90174-1}{ {Nucl. Phys.}
  {\bf B172} (1980) 413--424}.

\bibitem{Kaloper:2017fsa}
N.~Kaloper and J.~Terning, ``{{Landscaping the Strong CP Problem}},''
  \href{https://arxiv.org/abs/1710.01740}{arXiv:1710.01740 [hep-th]}.

\bibitem{Franco:2014hsa}
S.~Franco, D.~Galloni, A.~Retolaza, and A.~Uranga, ``{{On axion monodromy
  inflation in warped
  throats}},''\href{https://doi.org/10.1007/JHEP02(2015)086}{ {JHEP} {\bf 02}
  (2015) 086}, [\href{https://arxiv.org/abs/1405.7044}{arXiv:1405.7044
  [hep-th]}].

\bibitem{Bielleman:2015ina}
S.~Bielleman, L.~E. Ibanez, and I.~Valenzuela, ``{{Minkowski 3-forms, Flux
  String Vacua, Axion Stability and
  Naturalness}},''\href{https://doi.org/10.1007/JHEP12(2015)119}{ {JHEP} {\bf
  12} (2015) 119}, [\href{https://arxiv.org/abs/1507.06793}{arXiv:1507.06793
  [hep-th]}].

\bibitem{Valenzuela:2016yny}
I.~Valenzuela, ``{{Backreaction Issues in Axion Monodromy and Minkowski
  4-forms}},''\href{https://doi.org/10.1007/JHEP06(2017)098}{ {JHEP} {\bf 06}
  (2017) 098}, [\href{https://arxiv.org/abs/1611.00394}{arXiv:1611.00394
  [hep-th]}].

\bibitem{Montero:2017yja}
M.~Montero, A.~M. Uranga, and I.~Valenzuela, ``{{A Chern-Simons
  Pandemic}},''\href{https://doi.org/10.1007/JHEP07(2017)123}{ {JHEP} {\bf 07}
  (2017) 123}, [\href{https://arxiv.org/abs/1702.06147}{arXiv:1702.06147
  [hep-th]}].

\bibitem{Nitta:2018yzb}
M.~Nitta and R.~Yokokura, ``{{Higher derivative three-form gauge theories and
  their supersymmetric extension}},''{ {JHEP} {\bf 10} (2018) 146},
  [\href{https://arxiv.org/abs/1809.03957}{arXiv:1809.03957 [hep-th]}].

\bibitem{Ostrogradsky:1850fid}
M.~Ostrogradsky, ``{{M{\'e}moires sur les {\'e}quations diff{\'e}rentielles,
  relatives au probl{\`e}me des isop{\'e}rim{\`e}tres}},''{ {Mem. Acad. St.
  Petersbourg} {\bf 6} (1850) no.~4, 385--517}.

\bibitem{Woodard:2006nt}
R.~P. Woodard, ``{{Avoiding dark energy with 1/r modifications of
  gravity}},''\href{https://doi.org/10.1007/978-3-540-71013-4_14}{ {Lect. Notes
  Phys.} {\bf 720} (2007) 403--433},
  [\href{https://arxiv.org/abs/astro-ph/0601672}{arXiv:astro-ph/0601672
  [astro-ph]}].

\bibitem{Klinkhamer:2016jrt}
F.~R. Klinkhamer and G.~E. Volovik, ``{{Propagating q-field and q-ball
  solution}},''\href{https://doi.org/10.1142/S0217732317501036}{ {Mod. Phys.
  Lett.} {\bf A32} (2017) no.~18, 1750103},
  [\href{https://arxiv.org/abs/1609.03533}{arXiv:1609.03533 [hep-th]}].

\bibitem{Klinkhamer:2016zzh}
F.~R. Klinkhamer and G.~E. Volovik, ``{{Dark matter from dark energy in
  q-theory}},''\href{https://doi.org/10.1134/S0021364017020011}{ {JETP Lett.}
  {\bf 105} (2017) no.~2, 74--77},
  [\href{https://arxiv.org/abs/1612.02326}{arXiv:1612.02326 [physics.gen-ph]}].

\bibitem{Klinkhamer:2016lgk}
F.~R. Klinkhamer and G.~E. Volovik, ``{{More on cold dark matter from
  q-theory}},'' \href{https://arxiv.org/abs/1612.04235}{arXiv:1612.04235
  [gr-qc]}.

\bibitem{Klinkhamer:2017nfb}
F.~R. Klinkhamer and T.~Mistele, ``{{Classical stability of higher-derivative
  q-theory in the four-form-field-strength
  realization}},''\href{https://doi.org/10.1142/S0217751X17500907}{ {Int. J.
  Mod. Phys.} {\bf A32} (2017) no.~16, 1750090},
  [\href{https://arxiv.org/abs/1704.05436}{arXiv:1704.05436 [hep-th]}].

\bibitem{Khoury:2010gb}
J.~Khoury, J.-L. Lehners, and B.~Ovrut, ``{{Supersymmetric P(X,$\phi$) and the
  Ghost Condensate}},''\href{https://doi.org/10.1103/PhysRevD.83.125031}{
  {Phys. Rev.} {\bf D83} (2011) 125031},
  [\href{https://arxiv.org/abs/1012.3748}{arXiv:1012.3748 [hep-th]}].

\bibitem{Khoury:2011da}
J.~Khoury, J.-L. Lehners, and B.~A. Ovrut, ``{{Supersymmetric
  Galileons}},''\href{https://doi.org/10.1103/PhysRevD.84.043521}{ {Phys. Rev.}
  {\bf D84} (2011) 043521},
  [\href{https://arxiv.org/abs/1103.0003}{arXiv:1103.0003 [hep-th]}].

\bibitem{Koehn:2012te}
M.~Koehn, J.-L. Lehners, and B.~Ovrut, ``{{Ghost condensate in $N=1$
  supergravity}},''\href{https://doi.org/10.1103/PhysRevD.87.065022}{ {Phys.
  Rev.} {\bf D87} (2013) no.~6, 065022},
  [\href{https://arxiv.org/abs/1212.2185}{arXiv:1212.2185 [hep-th]}].

\bibitem{Nitta:2014pwa}
M.~Nitta and S.~Sasaki, ``{{BPS States in Supersymmetric Chiral Models with
  Higher Derivative
  Terms}},''\href{https://doi.org/10.1103/PhysRevD.90.105001}{ {Phys. Rev.}
  {\bf D90} (2014) no.~10, 105001},
  [\href{https://arxiv.org/abs/1406.7647}{arXiv:1406.7647 [hep-th]}].

\bibitem{Buchbinder:1994iw}
I.~L. Buchbinder, S.~Kuzenko, and Z.~Yarevskaya, ``{{Supersymmetric effective
  potential: Superfield
  approach}},''\href{https://doi.org/10.1016/0550-3213(94)90466-9}{ {Nucl.
  Phys.} {\bf B411} (1994) 665--692}.

\bibitem{Buchbinder:1994xq}
I.~L. Buchbinder, S.~M. Kuzenko, and A.~{\relax Yu}. Petrov, ``{{Superfield
  chiral effective
  potential}},''\href{https://doi.org/10.1016/0370-2693(94)90260-7}{ {Phys.
  Lett.} {\bf B321} (1994) 372--377}.

\bibitem{Banin:2006db}
A.~T. Banin, I.~L. Buchbinder, and N.~G. Pletnev, ``{{On quantum properties of
  the four-dimensional generic chiral superfield
  model}},''\href{https://doi.org/10.1103/PhysRevD.74.045010}{ {Phys. Rev.}
  {\bf D74} (2006) 045010},
  [\href{https://arxiv.org/abs/hep-th/0606242}{arXiv:hep-th/0606242 [hep-th]}].

\bibitem{Kuzenko:2014ypa}
S.~M. Kuzenko and S.~J. Tyler, ``{{The one-loop effective potential of the
  Wess-Zumino model
  revisited}},''\href{https://doi.org/10.1007/JHEP09(2014)135}{ {JHEP} {\bf 09}
  (2014) 135}, [\href{https://arxiv.org/abs/1407.5270}{arXiv:1407.5270
  [hep-th]}].

\bibitem{Koehn:2012ar}
M.~Koehn, J.-L. Lehners, and B.~A. Ovrut, ``{{Higher-Derivative Chiral
  Superfield Actions Coupled to N=1
  Supergravity}},''\href{https://doi.org/10.1103/PhysRevD.86.085019}{ {Phys.
  Rev.} {\bf D86} (2012) 085019},
  [\href{https://arxiv.org/abs/1207.3798}{arXiv:1207.3798 [hep-th]}].

\bibitem{Farakos:2012qu}
F.~Farakos and A.~Kehagias, ``{{Emerging Potentials in Higher-Derivative Gauged
  Chiral Models Coupled to N=1
  Supergravity}},''\href{https://doi.org/10.1007/JHEP11(2012)077}{ {JHEP} {\bf
  11} (2012) 077}, [\href{https://arxiv.org/abs/1207.4767}{arXiv:1207.4767
  [hep-th]}].

\bibitem{Queiruga:2016yzd}
J.~M. Queiruga, ``{{Supersymmetric galileons and auxiliary fields in 2+1
  dimensions}},''\href{https://doi.org/10.1103/PhysRevD.95.125001}{ {Phys.
  Rev.} {\bf D95} (2017) no.~12, 125001},
  [\href{https://arxiv.org/abs/1612.04727}{arXiv:1612.04727 [hep-th]}].

\bibitem{Sasaki:2012ka}
S.~Sasaki, M.~Yamaguchi, and D.~Yokoyama, ``{{Supersymmetric DBI
  inflation}},''\href{https://doi.org/10.1016/j.physletb.2012.10.006}{ {Phys.
  Lett.} {\bf B718} (2012) 1--4},
  [\href{https://arxiv.org/abs/1205.1353}{arXiv:1205.1353 [hep-th]}].

\bibitem{Aoki:2014pna}
S.~Aoki and Y.~Yamada, ``{{Inflation in supergravity without K{\"a}hler
  potential}},''\href{https://doi.org/10.1103/PhysRevD.90.127701}{ {Phys. Rev.}
  {\bf D90} (2014) no.~12, 127701},
  [\href{https://arxiv.org/abs/1409.4183}{arXiv:1409.4183 [hep-th]}].

\bibitem{Aoki:2015eba}
S.~Aoki and Y.~Yamada, ``{{Impacts of supersymmetric higher derivative terms on
  inflation models in
  supergravity}},''\href{https://doi.org/10.1088/1475-7516/2015/07/020}{ {JCAP}
  {\bf 1507} (2015) no.~07, 020},
  [\href{https://arxiv.org/abs/1504.07023}{arXiv:1504.07023 [hep-th]}].

\bibitem{Adam:2013awa}
C.~Adam, J.~M. Queiruga, J.~Sanchez-Guillen, and A.~Wereszczynski, ``{{Extended
  Supersymmetry and BPS solutions in baby Skyrme
  models}},''\href{https://doi.org/10.1007/JHEP05(2013)108}{ {JHEP} {\bf 05}
  (2013) 108}, [\href{https://arxiv.org/abs/1304.0774}{arXiv:1304.0774
  [hep-th]}].

\bibitem{Adam:2011hj}
C.~Adam, J.~M. Queiruga, J.~Sanchez-Guillen, and A.~Wereszczynski, ``{{N=1
  supersymmetric extension of the baby Skyrme
  model}},''\href{https://doi.org/10.1103/PhysRevD.84.025008}{ {Phys. Rev.}
  {\bf D84} (2011) 025008},
  [\href{https://arxiv.org/abs/1105.1168}{arXiv:1105.1168 [hep-th]}].

\bibitem{Nitta:2015uba}
M.~Nitta and S.~Sasaki, ``{{Classifying BPS States in Supersymmetric Gauge
  Theories Coupled to Higher Derivative Chiral
  Models}},''\href{https://doi.org/10.1103/PhysRevD.91.125025}{ {Phys. Rev.}
  {\bf D91} (2015) 125025},
  [\href{https://arxiv.org/abs/1504.08123}{arXiv:1504.08123 [hep-th]}].

\bibitem{Bolognesi:2014ova}
S.~Bolognesi and W.~Zakrzewski, ``{{Baby Skyrme Model, Near-BPS Approximations
  and Supersymmetric
  Extensions}},''\href{https://doi.org/10.1103/PhysRevD.91.045034}{ {Phys.
  Rev.} {\bf D91} (2015) no.~4, 045034},
  [\href{https://arxiv.org/abs/1407.3140}{arXiv:1407.3140 [hep-th]}].

\bibitem{Queiruga:2016jqu}
J.~M. Queiruga, ``{{Baby Skyrme model and fermionic zero
  modes}},''\href{https://doi.org/10.1103/PhysRevD.94.065022}{ {Phys. Rev.}
  {\bf D94} (2016) no.~6, 065022},
  [\href{https://arxiv.org/abs/1606.02869}{arXiv:1606.02869 [hep-th]}].

\bibitem{Queiruga:2018nph}
J.~M. Queiruga, ``{{SUSY Chern-Simons $\mathbb{CP}^N$ and baby Skyrme models
  and their BPS structures}},''\href{https://doi.org/10.1088/1751-8121/aaf93c}{
  {J. Phys.} {\bf A52} (2018) 055202},
  [\href{https://arxiv.org/abs/1807.09612}{arXiv:1807.09612 [hep-th]}].

\bibitem{Gudnason:2015ryh}
S.~B. Gudnason, M.~Nitta, and S.~Sasaki, ``{{A supersymmetric Skyrme
  model}},''\href{https://doi.org/10.1007/JHEP02(2016)074}{ {JHEP} {\bf 02}
  (2016) 074}, [\href{https://arxiv.org/abs/1512.07557}{arXiv:1512.07557
  [hep-th]}].

\bibitem{Gudnason:2016iex}
S.~B. Gudnason, M.~Nitta, and S.~Sasaki, ``{{Topological solitons in the
  supersymmetric Skyrme
  model}},''\href{https://doi.org/10.1007/JHEP01(2017)014}{ {JHEP} {\bf 01}
  (2017) 014}, [\href{https://arxiv.org/abs/1608.03526}{arXiv:1608.03526
  [hep-th]}].

\bibitem{Queiruga:2015xka}
J.~M. Queiruga, ``{{Skyrme-like models and supersymmetry in 3+1
  dimensions}},''\href{https://doi.org/10.1103/PhysRevD.92.105012}{ {Phys.
  Rev.} {\bf D92} (2015) no.~10, 105012},
  [\href{https://arxiv.org/abs/1508.06692}{arXiv:1508.06692 [hep-th]}].

\bibitem{Queiruga:2017blc}
J.~M. Queiruga and A.~Wereszczynski, ``{{Non-uniqueness of the supersymmetric
  extension of the $O(3)$
  $\sigma$-model}},''\href{https://doi.org/10.1007/JHEP11(2017)141}{ {JHEP}
  {\bf 11} (2017) 141},
  [\href{https://arxiv.org/abs/1703.07343}{arXiv:1703.07343 [hep-th]}].

\bibitem{Eto:2012qda}
M.~Eto, T.~Fujimori, M.~Nitta, K.~Ohashi, and N.~Sakai, ``{{Higher Derivative
  Corrections to Non-Abelian Vortex Effective
  Theory}},''\href{https://doi.org/10.1143/PTP.128.67}{ {Prog. Theor. Phys.}
  {\bf 128} (2012) 67--103},
  [\href{https://arxiv.org/abs/1204.0773}{arXiv:1204.0773 [hep-th]}].

\bibitem{Nitta:2014fca}
M.~Nitta and S.~Sasaki, ``{{Higher Derivative Corrections to Manifestly
  Supersymmetric Nonlinear
  Realizations}},''\href{https://doi.org/10.1103/PhysRevD.90.105002}{ {Phys.
  Rev.} {\bf D90} (2014) no.~10, 105002},
  [\href{https://arxiv.org/abs/1408.4210}{arXiv:1408.4210 [hep-th]}].

\bibitem{Nitta:2017yuf}
M.~Nitta, S.~Sasaki, and R.~Yokokura, ``{{Supersymmetry Breaking in Spatially
  Modulated Vacua}},''\href{https://doi.org/10.1103/PhysRevD.96.105022}{ {Phys.
  Rev.} {\bf D96} (2017) no.~10, 105022},
  [\href{https://arxiv.org/abs/1706.05232}{arXiv:1706.05232 [hep-th]}].

\bibitem{Nitta:2017mgk}
M.~Nitta, S.~Sasaki, and R.~Yokokura, ``{{Spatially Modulated Vacua in a
  Lorentz-invariant Scalar Field
  Theory}},''\href{https://doi.org/10.1140/epjc/s10052-018-6235-9}{ {Eur. Phys.
  J.} {\bf C78} (2018) no.~9, 754},
  [\href{https://arxiv.org/abs/1706.02938}{arXiv:1706.02938 [hep-th]}].

\bibitem{Farakos:2018sgq}
F.~Farakos, A.~Kehagias, and A.~Riotto, ``{{Liberated $ \mathcal{N} $ = 1
  supergravity}},''\href{https://doi.org/10.1007/JHEP06(2018)011}{ {JHEP} {\bf
  06} (2018) 011}, [\href{https://arxiv.org/abs/1805.01877}{arXiv:1805.01877
  [hep-th]}].

\bibitem{Cecotti:1986jy}
S.~Cecotti, S.~Ferrara, and L.~Girardello, ``{{Structure of the Scalar
  Potential in General $N=1$ Higher Derivative Supergravity in
  Four-dimensions}},''\href{https://doi.org/10.1016/0370-2693(87)91103-8}{
  {Phys. Lett.} {\bf B187} (1987) 321--326}.

\bibitem{Bagger:1996wp}
J.~Bagger and A.~Galperin, ``{{A New Goldstone multiplet for partially broken
  supersymmetry}},''\href{https://doi.org/10.1103/PhysRevD.55.1091}{ {Phys.
  Rev.} {\bf D55} (1997) 1091--1098},
  [\href{https://arxiv.org/abs/hep-th/9608177}{arXiv:hep-th/9608177 [hep-th]}].

\bibitem{Kuzenko:2000uh}
S.~M. Kuzenko and S.~Theisen, ``{{Nonlinear selfduality and
  supersymmetry}},''\href{https://doi.org/10.1002/1521-3978(200102)49:1/3<273::AID-PROP273>3.0.CO;2-0}{
  {Fortsch. Phys.} {\bf 49} (2001) 273--309},
  [\href{https://arxiv.org/abs/hep-th/0007231}{arXiv:hep-th/0007231 [hep-th]}].

\bibitem{Kuzenko:2002vk}
S.~M. Kuzenko and S.~A. McCarthy, ``{{Nonlinear selfduality and
  supergravity}},''\href{https://doi.org/10.1088/1126-6708/2003/02/038}{ {JHEP}
  {\bf 02} (2003) 038},
  [\href{https://arxiv.org/abs/hep-th/0212039}{arXiv:hep-th/0212039 [hep-th]}].

\bibitem{Antoniadis:2007xc}
I.~Antoniadis, E.~Dudas, and D.~M. Ghilencea, ``{{Supersymmetric Models with
  Higher Dimensional
  Operators}},''\href{https://doi.org/10.1088/1126-6708/2008/03/045}{ {JHEP}
  {\bf 03} (2008) 045}, [\href{https://arxiv.org/abs/0708.0383}{arXiv:0708.0383
  [hep-th]}].

\bibitem{Dudas:2015vka}
E.~Dudas and D.~M. Ghilencea, ``{{Effective operators in SUSY, superfield
  constraints and searches for a UV
  completion}},''\href{https://doi.org/10.1007/JHEP06(2015)124}{ {JHEP} {\bf
  06} (2015) 124}, [\href{https://arxiv.org/abs/1503.08319}{arXiv:1503.08319
  [hep-th]}].

\bibitem{Fujimori:2017kyi}
T.~Fujimori, M.~Nitta, K.~Ohashi, Y.~Yamada, and R.~Yokokura, ``{{Ghost-free
  vector superfield actions in supersymmetric higher-derivative
  theories}},''\href{https://doi.org/10.1007/JHEP09(2017)143}{ {JHEP} {\bf 09}
  (2017) 143}, [\href{https://arxiv.org/abs/1708.05129}{arXiv:1708.05129
  [hep-th]}].

\bibitem{Cribiori:2017laj}
N.~Cribiori, F.~Farakos, M.~Tournoy, and A.~van Proeyen, ``{{Fayet-Iliopoulos
  terms in supergravity without gauged
  R-symmetry}},''\href{https://doi.org/10.1007/JHEP04(2018)032}{ {JHEP} {\bf
  04} (2018) 032}, [\href{https://arxiv.org/abs/1712.08601}{arXiv:1712.08601
  [hep-th]}].

\bibitem{Aldabergenov:2018nzd}
Y.~Aldabergenov, S.~V. Ketov, and R.~Knoops, ``{{General couplings of a vector
  multiplet in $N=1$ supergravity with new FI
  terms}},''\href{https://doi.org/10.1016/j.physletb.2018.07.072}{ {Phys.
  Lett.} {\bf B785} (2018) 284--287},
  [\href{https://arxiv.org/abs/1806.04290}{arXiv:1806.04290 [hep-th]}].

\bibitem{Kuzenko:2018jlz}
S.~M. Kuzenko, ``{{Taking a vector supermultiplet apart: Alternative
  Fayet--Iliopoulos-type
  terms}},''\href{https://doi.org/10.1016/j.physletb.2018.04.051}{ {Phys.
  Lett.} {\bf B781} (2018) 723--727},
  [\href{https://arxiv.org/abs/1801.04794}{arXiv:1801.04794 [hep-th]}].

\bibitem{Aldabergenov:2017hvp}
Y.~Aldabergenov and S.~V. Ketov, ``{{Removing instability of inflation in
  Polonyi--Starobinsky supergravity by adding FI
  term}},''\href{https://doi.org/10.1142/S0217732318500323}{ {Mod. Phys. Lett.}
  {\bf A91} (2018) no.~05, 1850032},
  [\href{https://arxiv.org/abs/1711.06789}{arXiv:1711.06789 [hep-th]}].

\bibitem{Abe:2018plc}
H.~Abe, Y.~Aldabergenov, S.~Aoki, and S.~V. Ketov, ``{{Massive vector multiplet
  with Dirac-Born-Infeld and new Fayet-Iliopoulos terms in
  supergravity}},''\href{https://doi.org/10.1007/JHEP09(2018)094}{ {JHEP} {\bf
  09} (2018) 094}, [\href{https://arxiv.org/abs/1808.00669}{arXiv:1808.00669
  [hep-th]}].

\bibitem{Wess:1992cp}
J.~Wess and J.~Bagger, ``{{Supersymmetry and supergravity}},''{ {Princeton,
  USA: Univ. Pr.} (1992) }.

\bibitem{Cremmer:1982en}
E.~Cremmer, S.~Ferrara, L.~Girardello, and A.~Van~Proeyen, ``{{Yang-Mills
  Theories with Local Supersymmetry: Lagrangian, Transformation Laws and
  SuperHiggs Effect}},''\href{https://doi.org/10.1016/0550-3213(83)90679-X}{
  {Nucl. Phys.} {\bf B212} (1983) 413}.

\bibitem{Kugo:1982cu}
T.~Kugo and S.~Uehara, ``{{Conformal and Poincare Tensor Calculi in $N=1$
  Supergravity}},''\href{https://doi.org/10.1016/0550-3213(83)90463-7}{ {Nucl.
  Phys.} {\bf B226} (1983) 49--92}.

\bibitem{Kugo:1982mr}
T.~Kugo and S.~Uehara, ``{{Improved Superconformal Gauge Conditions in the
  $N=1$ Supergravity {Yang-Mills} Matter
  System}},''\href{https://doi.org/10.1016/0550-3213(83)90612-0}{ {Nucl. Phys.}
  {\bf B222} (1983) 125--138}.

\bibitem{Kugo:1983mv}
T.~Kugo and S.~Uehara, ``{{$N=1$ Superconformal Tensor Calculus: Multiplets
  With External Lorentz Indices and Spinor Derivative
  Operators}},''\href{https://doi.org/10.1143/PTP.73.235}{ {Prog. Theor. Phys.}
  {\bf 73} (1985) 235}.

\bibitem{Butter:2009cp}
D.~Butter, ``{{N=1 Conformal Superspace in Four
  Dimensions}},''\href{https://doi.org/10.1016/j.aop.2009.09.010}{ {Annals
  Phys.} {\bf 325} (2010) 1026--1080},
  [\href{https://arxiv.org/abs/0906.4399}{arXiv:0906.4399 [hep-th]}].

\bibitem{Kugo:2016zzf}
T.~Kugo, R.~Yokokura, and K.~Yoshioka, ``{{Component versus superspace
  approaches to D = 4, N = 1 conformal
  supergravity}},''\href{https://doi.org/10.1093/ptep/ptw090}{ {PTEP} {\bf
  2016} (2016) no.~7, 073B07},
  [\href{https://arxiv.org/abs/1602.04441}{arXiv:1602.04441 [hep-th]}].

\bibitem{Kugo:2016lum}
T.~Kugo, R.~Yokokura, and K.~Yoshioka, ``{{Superspace gauge fixing in
  Yang--Mills matter-coupled conformal
  supergravity}},''\href{https://doi.org/10.1093/ptep/ptw119}{ {PTEP} {\bf
  2016} (2016) no.~9, 093B03},
  [\href{https://arxiv.org/abs/1606.06515}{arXiv:1606.06515 [hep-th]}].

\end{thebibliography}\endgroup
\end{document}